\documentclass[12pt,onecolumn,draftcls]{IEEEtran}
\usepackage{verbatim}
\usepackage{graphicx}
\usepackage{amsmath,amssymb,amsfonts,bm}
\usepackage{subfigure}
\usepackage{psfrag}
\usepackage[dvips]{epsfig}
\usepackage{amsthm}
\usepackage{color}
\usepackage[usenames,dvipsnames]{pstricks}
\usepackage[dvips]{epsfig}
\usepackage{pst-grad} 
\usepackage{pst-plot} 
\usepackage{amsmath,graphicx}
\usepackage{amsmath,graphicx}
\usepackage{enumerate}
\usepackage{amsbsy}
\usepackage{amssymb}
\usepackage{amsthm}
\usepackage{amscd}
\usepackage{subfigure}
\usepackage{color}
\usepackage{cite}
\usepackage{listings}
\usepackage{tikz}
\usepackage{pgf}
\usepackage{stackrel}

\def\bx{{\bf x}}

\def\bb{{\bf b}}

\def\bI{{\bf I}}

\def\bA{{\bf A}}

\def\realR{{\mathbb{R}}}
\def\bx{{\bf x}}

\usepackage{amsmath}
\newcommand\addtag{\refstepcounter{equation}\tag{\theequation}}

\begin{document}
\title{Throughput Optimization for Wireless Powered Interference Channels}

\author{Omid Rezaei, Mohammad Mahdi Naghsh$^*$, \emph{Member, IEEE}, Zahra Rezaei, and Rui Zhang, \emph{Fellow, IEEE}

\thanks{O. Rezaei, M. M. Naghsh and Z. Rezaei are with the Department of Electrical and Computer Engineering, Isfahan University of Technology, Isfahan 84156-83111, Iran. R. Zhang is with the Department of Electrical and Computer Engineering, National University of Singapore, Singapore 117576.

*Please address all the correspondence to M. M. Naghsh, Phone: (+98) 31-33912450; Fax: (+98) 31-33912451; Email: mm\_naghsh@cc.iut.ac.ir} }

\maketitle

\begin{abstract}
This paper studies a general multi-user wireless powered interference channel (IFC) under the harvest-then-transmit protocol, where the communication in channel coherence time consists of two phases, namely wireless energy transfer (WET) and wireless information transfer (WIT). In the first phase, all energy transmitters (ETs) transmit energy signals to information transmitters (ITs) via collaborative waveform design, while in the second phase, each IT transmits an information signal to its intended ET using the harvested energy in the previous phase. The aim is to jointly design the WET-WIT time allocation, the (deterministic) transmit signal at the first phase, and the transmit power of ITs in the second phase to optimize the network throughput.  The design problems are non-convex and hence difficult to solve globally. To deal with them, we propose efficient iterative algorithms based on alternating projections; then, the majorization-minimization technique is used to tackle the non-convex sub-problems in each iteration. We also extend the devised design methodology by considering imperfect channel state information (CSI) and non-linearity in energy harvesting circuit. The proposed algorithms are locally convergent and can provide high-quality suboptimal solutions to the design problems. Simulation results show the effectiveness of the proposed algorithms under various setups.

\end{abstract}


\section{Introduction}
One of the major challenges in wireless networks is to prolong the lifetime of the conventional networks
which are powered by finite-capacity batteries. The network lifetime can be extended
by replacing/recharging the batteries; however, it might be expensive and even impossible especially in large-scale wireless network (e.g., wireless sensor networks) \cite{oppo}.
Therefore, energy harvesting (EH) technologies
have been considered as promising techniques to deal with this difficulty. The EH technologies are more user friendly/cost effective because they waive the need for manual battery charging/replacement.

The efficiency of the traditional and natural EH sources (such as solar, thermal, vibrational sources, etc.) highly depend upon time, location and the conditions of the environments. On the other hand, radio frequency (RF) enabled wireless energy transfer (WET) technology is a much more controllable and cost-efficient approach to prolong the lifetime of wireless networks \cite{oppo, wpcn, survey}. This technology provides wireless devices with continuous (harvested) energy from received RF signal instead of using conventional batteries. In this case, many practical advantages can be achieved including long operating range, simple and small harvester circuits, low production cost, and efficient energy multicasting \cite{oppo}. Indeed, due to the accumulative nature of EH, the interference signals received by an energy harvester can be a useful energy source in a wireless communication network.

\subsection{Related Works} \label{i1}
In the literature, there are two lines of research for WET-based communications: wireless powered communication network (WPCN) and simultaneous wireless information and power transfer (SWIPT). In WPCNs, wireless devices are powered by dedicated WET in the downlink in order to transfer information in the uplink; whereas in SWIPTs, a dual use of RF signals is considered for simultaneous WET and wireless information transfer (WIT) \cite{onebit}. The design of WPCNs and SWIPTs for different setups has been addressed in numerous works (e.g., \cite{Throughput, sdma, full2, placement, hina2, large, enable,park2013joint,coll, power, IF-CH, zhu}). Particularly, in \cite{Throughput}, a “harvest-then-transmit” protocol was proposed for a multi-user WPCN, where users first harvest energy from RF signals which is broadcasted by a single-antenna hybrid access point (HAP) in the downlink. Then, they transmit independent information to the HAP in the uplink via time-division-multiple-access (TDMA) using the harvested energy. In this work, the downlink WET time slot and uplink information transmission time slots for all users have been jointly optimized to maximize the network throughput. The authors in \cite{sdma} have extended \cite{Throughput} to a multi-antenna WPCN scenario, where a multi-antenna HAP enables the uplink transmission via space-division-multiple-access. The reference \cite{park2013joint} considered interference channel (IFC) under SWIPT setting; the work in \cite{coll} proposed collaborative energy beamforming (EB) with distributed single-antenna transmitters for SWIPT under an IFC setup.

 Note that there are also prior works in the literature that focus on the waveform design in wireless power transfer (WPT) and SWIPT to maximize the amount of harvested power and thus enhance the WPT efficiency \cite{clerckx2016waveform, huang2018waveform, clerckx2017wireless}. For example, the authors in \cite{clerckx2016waveform} considered the optimum multi-sine waveform design for WPT systems to maximize the output current of the energy receiver circuit. Note that in the the above works, the amount of harvested energy is random due to the randomness of the energy/information signals; this randomness leads to an uncertainty of the instantaneous harvested energy. This issue can be resolved by assuming deterministic energy signals in WPCN (to be considered in this paper). Imperfect channel state information (CSI) has been taken into account in \cite{ng2014robust, zhou2017robust, feng2015robust, brobust} for SWIPT and WPCN. In \cite{brobust}, the authors studied a robust resource allocation in a TDMA-based MIMO-WPCN with a non-linear EH model. The
non-linearity has also been considered in \cite{clerckx2017wireless} and \cite{boshkovska2015practical} for SWIPT.

\subsection{Contributions}\label{i2}
In this paper, we consider an IFC adopting the harvest-then-transmit protocol with multiple transmit-receive pairs. Similar to conventional IFC, information transmitters (ITs) transmit their independent information signals to their intended receivers; however, the difference lies in that we consider  ITs have no conventional energy supplies. Therefore, they harvest the energy from information receivers (IRs) in advance. Specifically , there are two phases in this setup. In the first phase, all IRs act as energy transmitters (ETs) and transmit deterministic energy signals\footnote[1]{We consider deterministic energy signals for energy transmission in phase 1 (see \cite{clerckx2017wireless} for a similar assumption). By doing so, the amount of the harvested energy is not random and can be reliably used for WIT in  the second phase. On the other hand, when using random energy signals, the interfering signals may be combined destructively/constructively at the energy receivers in the first phase, leading to considerable instantaneous variations of the harvested energy.} with collaborative waveform design. In the second phase, each energy receiver (ER) transmits its own information signal to its intended IR using the harvested energy in the previous phase and acts as an IT\footnote[2]{Note that, an example of the considered scenario is the case with one HAP (ET/IR) that serves a single tier of users (ER/IT).}. The main contributions of this work are summarized as follows:
\begin{itemize}
\item

We consider a $\mathcal{K}$-link WPCN where all devices work in the same frequency band, i.e., the uplink WIT can be modeled as an IFC . We also extend the work by considering imperfect CSI and non-linearity in EH circuit.

\item

We aim to optimize network throughputs in both \emph{sum} and \emph{max-min} senses. The resulted design problems are non-convex and hence, hard to solve. Therefore, we devise a method based on alternating projections in order to solve the problems suboptimally but efficiently. The resulted subproblems are still non-convex and thus, we resort to the majorization-minimization technique to handle them efficiently.

\item

In addition to designing power control in the second phase and the time division parameter, we also consider the waveform design methodology to design the deterministic transmit signal in the energy transmission phase (i.e., energy waveform).

\end{itemize}

\subsection{Organization} \label{i3}
The rest of this paper is organized as follows. In Section \ref{sys}, we present the IFC with harvest-then-transmit protocol and the problem formulation. In Section \ref{sum}, the sum throughput maximization problem is studied and the proposed method for solving it is presented. In Section \ref{maxmin}, we formulate the max-min throughput optimization problem and propose an algorithm to solve it. Numerical results are provided in Section \ref{simu} and finally the
conclusions are drawn in Section \ref{con}.
\begin{figure}
\centering
\begin{tikzpicture}[even odd rule,rounded corners=2pt,x=12pt,y=12pt,scale=.9]
\draw[thick,fill=green!10] (-15,5) rectangle ++(2.5,2) node[midway]{$\textrm{ET}_1$};
\draw[thick] (-12.5,6)--++(1.5,0)--+(0,1);
\draw[thick,fill=gray!15] (-11,7)--++(.75,0.5)--++(-1.5,0)--++(.75,-.5)--++(0,-.1);
\draw (-11,4.5) node {\huge$\vdots$};
\draw[thick,fill=green!10] (-15,0) rectangle ++(2.5,2) node[midway]{ $\textrm{ET}_k$};
\draw[thick] (-12.5,1)--++(1.5,0)--+(0,1);
\draw[thick,fill=gray!15] (-11,2)--++(.75,0.5)--++(-1.5,0)--++(.75,-.5)--++(0,-.1);
\draw (-11,-.5) node {\huge$\vdots$};
\draw[->,line width=.5mm,green!100] (-10,2.4)--++(6.8,4.9);
\draw[->,line width=.5mm,green!100] (-10,2.4)--++(6.8,-4.9);
\draw[->,line width=.5mm,green!100] (-10,2.4)--++(6.8,0);
\draw[thick,fill=green!10] (-15,-5) rectangle ++(2.5,2) node[midway]{$\textrm{ET}_\mathcal{K}$};
\draw[thick] (-12.5,-4)--++(1.5,0)--+(0,1);
\draw[thick,fill=gray!15] (-11,-3)--++(.75,0.5)--++(-1.5,0)--++(.75,-.5)--++(0,-.1);
\draw[thick,fill=red!10] (-.5,5) rectangle ++(2.5,2) node[midway]{ $\textrm{ER}_1$};
\draw[thick] (-.5,6)--++(-1.5,0)--+(0,1);
\draw[thick,fill=gray!15] (-2,7)--++(.75,0.5)--++(-1.5,0)--++(.75,-.5)--++(0,-.1);
\draw (-2,4.5) node {\huge$\vdots$};
\draw[thick,fill=red!10] (-.5,0) rectangle ++(2.5,2) node[midway]{ $\textrm{ER}_k$};
\draw[thick] (-.5,1)--++(-1.5,0)--+(0,1);
\draw[thick,fill=gray!15] (-2,2)--++(.75,0.5)--++(-1.5,0)--++(.75,-.5)--++(0,-.1);
\draw (-2,-.5) node {\huge$\vdots$};
\draw[thick,fill=red!10] (-.5,-5) rectangle ++(2.5,2) node[midway]{ $\textrm{ER}_\mathcal{K}$};
\draw[thick] (-.5,-4)--++(-1.5,0)--+(0,1);
\draw[thick,fill=gray!15] (-2,-3)--++(.75,0.5)--++(-1.5,0)--++(.75,-.5)--++(0,-.1);
\draw[thick,fill=green!10] (5,5) rectangle ++(2.5,2) node[midway]{$\textrm{IR}_1$};
\draw[thick] (7.5,6)--++(1.5,0)--+(0,1);
\draw[thick,fill=gray!15] (9,7)--++(.75,0.5)--++(-1.5,0)--++(.75,-.5)--++(0,-.1);
\draw (9,4.5) node {\huge$\vdots$};
\draw[thick,fill=green!10] (5,0) rectangle ++(2.5,2) node[midway]{ $\textrm{IR}_k$};
\draw[thick] (7.5,1)--++(1.5,0)--+(0,1);
\draw[thick,fill=gray!15] (9,2)--++(.75,0.5)--++(-1.5,0)--++(.75,-.5)--++(0,-.1);
\draw (9,-.5) node {\huge$\vdots$};
\draw[thick,fill=green!10] (5,-5) rectangle ++(2.5,2) node[midway]{$\textrm{IR}_\mathcal{K}$};
\draw[thick] (7.5,-4)--++(1.5,0)--+(0,1);
\draw[thick,fill=gray!15] (9,-3)--++(.75,0.5)--++(-1.5,0)--++(.75,-.5)--++(0,-.1);
\draw[thick,fill=red!10] (19.5,5) rectangle ++(2.5,2) node[midway]{ $\textrm{IT}_1$};
\draw[thick] (19.5,6)--++(-1.5,0)--+(0,1);
\draw[thick,fill=gray!15] (18,7)--++(.75,0.5)--++(-1.5,0)--++(.75,-.5)--++(0,-.1);
\draw (18,4.5) node {\huge$\vdots$};
\draw[thick,fill=red!10] (19.5,0) rectangle ++(2.5,2) node[midway]{ $\textrm{IT}_k$};
\draw[thick] (19.5,1)--++(-1.5,0)--+(0,1);
\draw[thick,fill=gray!15] (18,2)--++(.75,0.5)--++(-1.5,0)--++(.75,-.5)--++(0,-.1);
\draw (18,-.5) node {\huge$\vdots$};
\draw[->,line width=.5mm,red!100] (17,2.4)--++(-6.8,4.9);
\draw[->,line width=.5mm,red!100] (17,2.4)--++(-6.8,-4.9);
\draw[->,line width=.5mm,black!100] (17,2.4)--++(-6.8,0);
\draw[thick,fill=red!10] (19.5,-5) rectangle ++(2.5,2) node[midway]{ $\textrm{IT}_\mathcal{K}$};
\draw[thick] (19.5,-4)--++(-1.5,0)--+(0,1);
\draw[thick,fill=gray!15] (18,-3)--++(.75,0.5)--++(-1.5,0)--++(.75,-.5)--++(0,-.1);
\node [] at (-6.5,-7) {phase 1: energy transfer};
\node [] at (-6.5,-8.6) {$0 < t < \tau$};
\node [] at (13.7,-7) {phase 2: information transfer};
\node [] at (13.7,-8.6) {$\tau < t < T$};
\draw[->,line width=.5mm,black!100] (-8.8,-12)--+(3.9,0);
\draw[->,line width=.5mm,red!100] (-10.4,-13.1)--+(5.4,0);
\draw[->,line width=.5mm,green!100] (-10.35,-11)--+(5.4,0);
\node [] at (-13,-11) {\footnotesize energy flow};
\node [] at (-12.1,-12) {\footnotesize information flow};
\node [] at (-13,-13) {\footnotesize interference};
\end{tikzpicture}
\caption{A $\mathcal{K}$-link IFC with harvest-then-transmit protocol. In phase 1, all ETs transmit their energy signals to all ERs with collaboration. Due to the accumulative nature of EH, each ER harvests energy from all received energy signals. In phase 2, all ITs send their information signals to their intended IRs using the harvested energy in the previous phase simultaneously (where an ER in phase 1 act as an IT in phase 2). Note that in phase 2, the information signals generated by other ITs cause co-channel interference that is harmful for the WIT.}
\label{ht}
\centering
\end{figure}
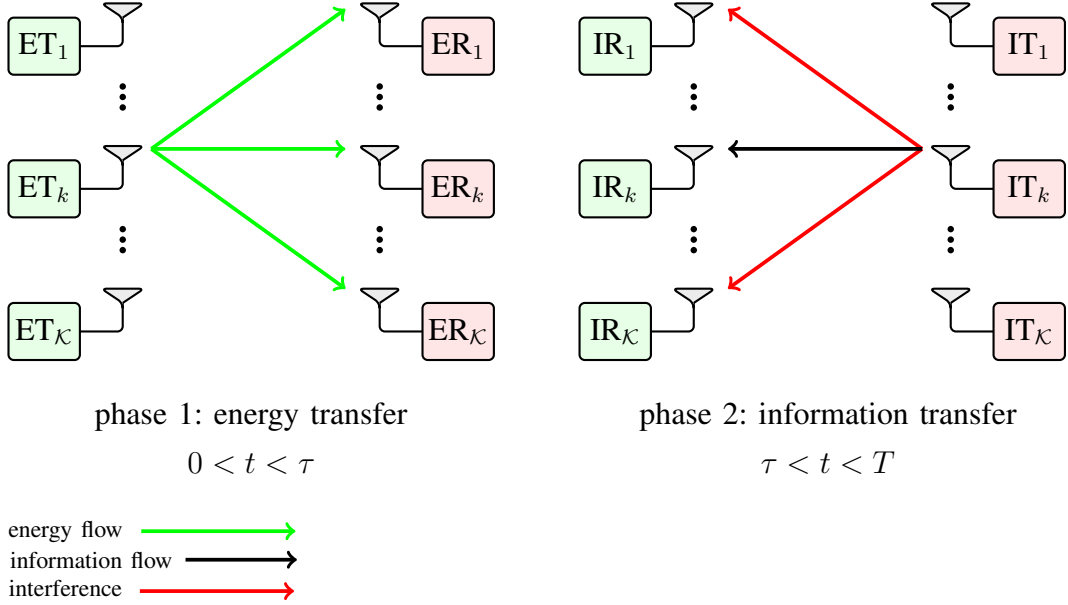

\emph{Notation:}  Bold lowercase letters and bold uppercase letters are used for vectors and matrices respectively. $\bI_{N}$ represents the identity matrix. We denote vector/matrix transpose by ${(\cdot)^T}$, the Hermitian by $(\cdot)^H$ and the complex conjugate by $(\cdot)^*$. Notations $\nabla {{{f}} (\mathbf{x}) }$ and ${\nabla}^{2} {{{f}} (\mathbf{x}) }$ denote the gradient and the Hessian of the twice-differentiable function ${{{f}} (\mathbf{x}) }$, respectively. $\Re \lbrace x \rbrace$ represents the real part of a complex number $x$ and $\textrm{arg}(x)$ denotes its phase argument. The notation $\mathbf{\lambda}_{\textrm{max}}(\cdot)$ indicates the maximum (principal) eigenvalue of a matrix. The $l_2$-norm of a vector $\bx$ is denoted by $\|\bx\|_2$. $\mbox{tr}\lbrace\cdot\rbrace$ denotes the trace of a square matrix and $\mathbb{E}\{\cdot\}$ stands for the statistical expectation. Finally, the notation  $\bA \succeq \mathbf{0}$ implies that the matrix $\bA$ is positive semidefinite.

\section{System Model} \label{sys}
We consider a wireless powered ${\cal K}$-user IFC with single-antenna nodes in which ITs have no conventional energy supplies (e.g., fixed batteries) and thus they need to replenish energy from the signals sent by the ETs in the network. 

In the sequel, we adopt the “harvest-then-transmit” protocol proposed in \cite{Throughput}, as shown
in Fig. \ref{ht}. In each block of duration $T$, during
the first phase (i.e. time slot $ t \in [ 0 , \tau ]$), all ETs collaboratively broadcast
energy signals to all ITs simultaneously. Hence, in phase 1, ITs act as ERs. In phase 2 (i.e. time slot $ t \in [ \tau , T]$), ITs transmit their own information to the associated IRs simultaneously. This is performed using the harvested energy during phase 1. Note that, an IR in phase 2 is also an ET in phase 1. As mentioned earlier, in the considered model, an ET/IR can be a HAP that serves a single-tier of users (ER/IT) (see also \cite{coll} and \cite{power} for similar scenarios).


The transmit power of the $k^{th}$ ET (i.e., $\textrm{ET}_k$) for $k = 1, . . . , \mathcal{K}$ is limited by the maximum power budget which is denoted by $p_{\textrm{max},k}$.
The channel from the $k^{th}$ ET to the $j^{th}$ IT and the corresponding reverse channel are denoted by (complex) random variables $h_{j,k}$ and $g_{j,k}$, respectively, with gains $\vert h_{j,k} \vert^2$ and $\vert g_{j,k} \vert^2$.
Let the deterministic transmitted (baseband) signal of the $k^{th}$ ET at phase 1 be $ x_k $ and define the corresponding transmit signal $\mathbf{x} = [x_1 ,x_2 , ... ,x_{\mathcal{K}}]^{T}$ as the energy waveform\footnote[1]{Herein, an option for the continuous-time transmit signal of the $k^{th}$ ET, i.e. $s_k(t)$, is $s_k(t) = \Re \lbrace x_k \textrm{exp}(j \omega_c t) \rbrace$. However, multi-carrier deterministic signals, modulated signals with Circularly Symmetric Complex Gaussian (CSCG) inputs or flash-based signaling can lead to higher harvested power (see \cite{clerckx2016waveform,clerckx2017wireless,varasteh2017capacity}).}.

As we adopt collaborative energy waveform design, the received signal at the $k^{th}$ ER at this phase is given by
\begin{equation}
y_k = \mathbf{h}^{H}_{k} \mathbf{x}  , \hspace{8pt} \forall k,
\label{Q10}
\end{equation}
with $\mathbf{h}_{k}\triangleq{[{h}^{*}_{k,1} ,{h}^{*}_{k,2}, ... ,{h}^{*}_{k,\mathcal{K}}]}^{T}$. Note that according to the IFC model, the $k^{th}$ ER can harvest the wireless energy from not only the $k^{th}$ ET but also other ETs' signals. Therefore, given channel coefficient ${\mathbf{h}}_{k}$, the amount of harvested energy by the $k^{th}$ ER is given by 
\begin{align} \label{Q}
E_k (\mathbf{x})&=\mu_k \tau {\left \vert  y_k \right \vert}^2  =\mu_k \tau  \mathbf{h}^{H}_{k} \mathbf{x} {\mathbf{x}}^{H} {\mathbf{h}}_{k}, \hspace{8pt} \forall k,
\end{align}
with $\mu_k$ being a constant associated with the linear EH circuit assumed. After the ERs replenish their energy during phase 1, they transmit (independent) information to their associated receivers (IRs) in phase 2. In fact, the energy of the information signal in this WIT phase is limited by the sum of the harvested energy in the WET phase of the current period and remaining energy from previous periods ($E_{k} (\mathbf{x})+ E_{0,k}$). Precisely, the transmit power for the information transmission at phase 2 (${p}_k $) has a constraint as follows
\begin{equation}
p_{{c}_k} + {{\varepsilon}_k}(T- \tau) {{p}}_{k} \leq E_{k} (\mathbf{x}) + E_{0,k}  , \hspace{8pt} \forall k,
\end{equation}
where $p_{{c}_k}$ is the constant related to circuit power consumption, ${{\varepsilon}_k}$ stands for power amplifier
efficiency with $0<({\varepsilon}_k) \leq1$, $E_{0,k}$ is the remaining energy of the previous period, and ${p}_k$ is the design parameter to be optimized.
In the following, without loss of generality, we consider a normalized time duration $T=1$.
%

The received signal $w_k$ at the $k^{th}$ IR in phase 2 can be expressed as
\begin{equation}
w_k = g_{k,k} \sqrt{{p}_k}{s}_k + \sum_{j=1 , j\neq k}^{\mathcal{K}} g_{j,k} \sqrt{{p}_j}{s}_j + n_k, \hspace{8pt} \forall k,
\label{sin2r}
\end{equation}
where ${s}_k$ is the information symbol from the $k^{th}$ transmitter at phase 2 and $n_k$ denotes the zero-mean additive white Gaussian noise (AWGN) at the $k^{th}$ IR with variance $\sigma_k^2$.
Consequently, assuming availability of perfect CSI \cite{Throughput}, the signal-to-interference-plus-noise ratio (SINR) associated with the $k^{th}$ ET-IT pair is given by
\begin{equation} \label{sin10r}
\gamma_k ({\mathbf{p}}) =\frac{ \vert {g}_{k,k} \vert ^2 {{p}}_k}{ \sum_{j=1, j\neq k}^{\mathcal{K}} \vert {g}_{j,k} \vert ^2 {{p}}_{j}+ \sigma _k^2}, \hspace{8pt} \forall k,
\end{equation}
with $ {\mathbf{p}}\triangleq [ {{p}}_1, {{p}}_2, ..., {{p}}_{\mathcal{K}}]^T$.
In light of the above expression for the SINR, the achievable throughput  bits/second/Hertz (bps/Hz) associated with the $k^{th}$ ET-IT pair (at phase 2) can be expressed as
\begin{equation}
R_k ({\mathbf{p}}, \tau) = (1 - \tau) \textrm{log}_2 (1 + \gamma_k ({\mathbf{p}})), \hspace{8pt} \forall k.
\label{ratek}
\end{equation}
In the next sections, we consider different throughput optimization problems with respect to (w.r.t.) the WET-WIT time allocation (i.e., the parameter $\tau$), the transmit waveform in phase 1 (i.e., $\mathbf{x}$), and the transmit power in phase 2 (i.e., ${\mathbf{p}}$) based on the above system model. Specifically, we deal with the sum- and min-throughput maximization problems in Sections III and IV, respectively.
\section{Sum Throughput Maximization}\label{sum}
In this section, we aim to maximize the sum throughput of the network, i.e., $\sum_{k=1}^{\mathcal{K}} R_k ({\mathbf{p}} ,\tau)$. According to \eqref{ratek}, the sum throughput maximization problem can be cast as
\begin{eqnarray}\label{form3}
\max_{\mathbf{x},{\mathbf{p}} ,\tau} && (1 - \tau)\sum_{k=1}^{\mathcal{K}} \textrm{log}_2 (1 + \gamma_k ({\mathbf{p}})) \\ \nonumber
\mbox{s. t.}\;\;&&
\textrm{C}_{1}: 0 \leq \tau \leq 1, \\ \nonumber &&
\textrm{C}_{2}: {\vert {x}_{k} \vert}^{2} \leq {p}_{\textrm{max},{k}}, \forall k, \\ \nonumber
&& \textrm{C}_{3}: p_{{c}_k} + {{\varepsilon}_k}(1- \tau) {{p}}_{k} \leq E_{k}(\mathbf{x})+ E_{0,k}, \forall k,\\ \nonumber
&& \textrm{C}_{4}: E_{k}(\mathbf{x}) + E_{0,k} \leq E_{{\textrm{max}},{k}}, \forall k.
\end{eqnarray}
Note that constraint $\textrm{C}_{4}$ states that each ER has a finite capacity for energy storage $E_{{\textrm{max}},{k}}$ \cite{lee2016resource,cl1,cl2}.
Using \eqref{Q} and \eqref{sin10r}, this problem can be rewritten as
\begin{eqnarray}\label{form30}
\max_{\mathbf{x},{\mathbf{p}} ,\tau} && (1 - \tau)\sum_{k=1}^{\mathcal{K}} \textrm{log}_2 \left(1 + \frac{ \vert {g}_{k,k} \vert ^2 {{p}}_k}{ \sum_{j=1, k\neq j}^{\mathcal{K}} \vert {g}_{j,k} \vert ^2 {{p}}_{j}+ \sigma _k^2} \right) \\ \nonumber
\mbox{s. t.}\;\;&&
\textrm{C}_{1}, \textrm{C}_{2}, \\ \nonumber
&& \textrm{C}_{3}: p_{{c}_k} + {{\varepsilon}_k}(1- \tau) {{p}}_{k} \leq \mu_k \tau {\mathbf{x}}^{H}  {{\mathbf{h}}}_{k} {{{\mathbf{h}}}}^{H}_{k}    \mathbf{x} + E_{0,k}, \forall k,\\ \nonumber
&& \textrm{C}_{4}: \mu_k \tau {\mathbf{x}}^{H}  {{\mathbf{h}}}_{k} {{{\mathbf{h}}}}^{H}_{k}    \mathbf{x}+ E_{0,k} \leq E_{{\textrm{max}},{k}}, \forall k.
\end{eqnarray}
The problem in \eqref{form30} is non-convex due to the coupled design variables in the objective function and the constraint set.
In the sequel, we employ the alternating projections approach \cite{stoica2004cyclic} by partitioning $[ \mathbf{x} , {\mathbf{p}} ]$ and $\tau$ to deal with this problem. More specifically, we first consider the problem w.r.t. $\mathbf{x}$ and ${\mathbf{p}}$ for fixed $\tau$ and then w.r.t. $\tau$ for fixed $\mathbf{x}$ and ${\mathbf{p}}$; the procedure continues till a stop criterion is satisfied (See Remark 1). As the aforementioned optimizations with certain fixed variables are still non-convex, we need to tackle them via the majorization-minimization technique.
\subsection{Optimizing $\mathbf{x}$ and ${\mathbf{p}}$ for fixed $\tau$} \label{suma}
Let
\begin{equation*}
a_{k} \triangleq \vert {g}_{k,k} \vert ^2 , \hspace{.2cm} \forall k, \hspace{.5cm}
b_{k,j} \triangleq
\vert {g}_{j,k} \vert ^2 , \hspace{.2cm}\forall k \neq j,
\end{equation*}
and note that the SINR associated with the $k^{th}$ pair in \eqref{sin10r} can be rewritten as the following expression
\begin{equation} \label{tr}
\gamma_k ( {\mathbf{p}}) = \frac{ { \mathbf{a}}_k^{T}{\mathbf{p}}}{{ \mathbf{b}}_k^{T}{\mathbf{p}}+ \sigma _k^2},
\end{equation}
with ${\mathbf{a}}_k\triangleq {a}_{k} \mathbf{e_k}$ where $\mathbf{e_k}$ is the $k^{th}$ standard vector and ${\mathbf{b}}_k\triangleq [{b}_{k,1}, {b}_{k,2}, ... , {b}_{k,k-1},0, {b}_{k,k+1}, ... , {b}_{k,\mathcal{K}}]^T$. The problem in \eqref{form30} for fixed $\tau$ reduces to the following optimization problem:
\begin{eqnarray}\label{sum1}
\max_{\mathbf{x},{\mathbf{p}} } && (1 - \tau){\sum_{k=1}^{\mathcal{K}}\textrm{log}_2 \left(1+ \frac{ { \mathbf{a}}_k^{T}{\mathbf{p}}}{{ \mathbf{b}}_k^{T}{\mathbf{p}}+ \sigma _k^2}\right)} \\ \nonumber
\mbox{s. t.}\;\;&& \textrm{C}_{2}, \textrm{C}_{3}, \textrm{C}_{4}.
\end{eqnarray}
This problem is non-convex w.r.t. $[ \mathbf{x} , {\mathbf{p}} ]$ due to the non-concave objective function and the non-convex constraint $\textrm{C}_{3}$. Therefore, we resort to the majorization-minimization (MaMi) technique\footnote[1]{Also known as the minorization-maximization (MM) technique. } to deal with the problem.
MaMi is an iterative method that can be used to obtain a suboptimal solution to any non-convex optimization problem in the general form of :
\begin{align}\label{mami1}
P_0: \quad \left\{
\begin{array}{ll}
&\underset{\widetilde{\bx}} {\max}\;\; f(\widetilde{\bx}) \\
& \textrm{s.t.} \quad \widetilde{g}(\widetilde{\bx})\leq 0.
\end{array}
\right.
\end{align}
where, $f(\widetilde{\bx})$ and $\widetilde{g}(\widetilde{\bx})$ can be non-concave and non-convex functions, respectively. To apply MaMi to $P_0$, we should obtain two functions at the $i^{th}$ iteration, namely  $h^{(i)}(\widetilde{\bx})$ and $q^{(i)}(\widetilde{\bx})$, such that $q^{(i)}(\widetilde{\bx})$ minorizes $f(\widetilde{\bx})$, i.e.,
\begin{align}\label{mami3}
&f(\widetilde{\bx}) \geq q^{(i)}(\widetilde{\bx}) , \quad \forall \widetilde{\bx}, \\ \nonumber
&f({\widetilde{\bx}}^{(i-1)})=q^{(i)}({\widetilde{\bx}}^{(i-1)}),
\end{align}
and $h^{(i)}(\widetilde{\bx})$ majorizes $\widetilde{g}(\widetilde{\bx})$, i.e., 
\begin{align}\label{mami2}
&h^{(i)}(\widetilde{\bx})\geq \widetilde{g}(\widetilde{\bx}), \quad \forall \widetilde{\bx}, \\ \nonumber
&h^{(i)}({\widetilde{\bx}}^{(i-1)})=\widetilde{g}({\widetilde{\bx}}^{(i-1)}),
\end{align}
where ${\widetilde{\bx}}^{(i-1)}$ is the value of $\widetilde{\bx}$ at the $(i-1)^{th}$ iteration. Next, the following optimization problem is solved at the $i^{th}$ iteration (which is simpler than the original problem):
\begin{align}\label{mami33}
P_i: \quad \left\{
\begin{array}{ll}
&\underset{\widetilde{\bx}} {\max}\;\; q^{(i)}(\widetilde{\bx}) \\
& \textrm{s.t.} \quad h^{(i)}(\widetilde{\bx})\leq 0.
\end{array}
\right.
\end{align}

Let $ \{\mathbf{q}_k\}_k\triangleq \{\mathbf{a} _k+\mathbf{b} _k\}_k$ and note that the optimization in \eqref{sum1} can be recast as
\begin{eqnarray}\label{sum2}
\max_{\mathbf{x},{\mathbf{p}} } && (1 - \tau){\sum_{k=1}^{\mathcal{K}}\textrm{log}_2 \left( \frac{ {\mathbf{q}}_k^{T}{\mathbf{p}}+\sigma _k^2}{ {\mathbf{b}}_k^{T} {\mathbf{p}} +\sigma _k^2}\right)} \\ \nonumber
\mbox{s. t.}\;\;&& \textrm{C}_{2}, \textrm{C}_{3}, \textrm{C}_{4}.
\end{eqnarray}
The problem in (\ref{sum2}) can be equivalently expressed as the following problem
\begin{eqnarray}\label{sum20}
\max_{\mathbf{x},{\mathbf{p}} } && (1-\tau)\displaystyle \sum_{k=1}^{\mathcal{K}} \left\lbrace \textrm{log}_2 ( \mathbf{q}_k^{T} {\mathbf{p}}+\sigma _k^2) - \textrm{log}_2 ( \mathbf{b}_k^{T} {\mathbf{p}}+\sigma _k^2) \right\rbrace \\ \nonumber
\mbox{s. t.}\;\;&& \textrm{C}_{2}, \textrm{C}_{3}, \textrm{C}_{4}.
\end{eqnarray}
Next, let $f_{1,k} ({\mathbf{p}}) \triangleq \textrm{log}_2 ( \mathbf{q}_k^{T} {\mathbf{p}}+\sigma _k^2) $ and $f_{2,k} ({\mathbf{p}}) \triangleq - \textrm{log}_2 ( \mathbf{b}_k^{T} {\mathbf{p}}+\sigma _k^2)$. Now, $f_{1,k}({\mathbf{p}})$ and $f_{2,k}({\mathbf{p}})$ are concave and convex functions of ${\mathbf{p}}$, respectively.
Note that for the problem in (\ref{sum20}), the objective function and the constraint $\textrm{C}_{3}$ are non-concave/convex w.r.t. ${\mathbf{p}}$ and ${\mathbf{x}}$, respectively. Accordingly, we start by dealing with the non-concave  objective function in \eqref{sum20} via MaMi.
To apply MaMi to the objective in \eqref{sum20}, we should minorize $\left \lbrace \sum_{k=1}^{\mathcal{K}} \left \lbrace f_{1,k}({\mathbf{p}})+f_{2,k}({\mathbf{p}}) \right \rbrace \right \rbrace$. To this end, we keep the function $f_{1,k}({\mathbf{p}})$ and minorize $f_{2,k}({\mathbf{p}})$, for every $k$. To obtain the minorizer, we observe the following inequality (which is concluded from the concavity of the function $\log (t)$ for $t \in \realR_+$):
\begin{equation}\label{ineq}
\textrm{log} (t) \leq \textrm{log} (t_0) + \frac{1}{t_0} (t - t_0).
\end{equation}
Next, note that setting $t \triangleq \mathbf{ b}_k^{T} {\mathbf{p}} +\sigma _k^2$ leads to the below minorizer for $f_2({\mathbf{p}})$:
\begin{equation} \label{new3}
- \textrm{log}_2 ( \mathbf{ b}_k^{T}{\mathbf{p}}+\sigma _k^2) \geq - \textrm{log}_2 ( \mathbf{ b}_k^{T}{{\mathbf{p}}}_0 +\sigma _k^2)-\frac{1}{ \mathbf{b}_k^{T}{{\mathbf{p}}}_0+\sigma _k^2} ( \mathbf{ b}_k^{T} {\mathbf{p}} - \mathbf{ b}_k^{T}{{\mathbf{p}}}_0).
\end{equation}
By substituting the above minorizer in the objective of \eqref{sum20} and neglecting the constant terms, the following objective is obtained at the $i^{th}$ iteration of the MaMi technique:
\begin{eqnarray}
\displaystyle \max_{\mathbf{x},{\mathbf{p}} } && (1-\tau)\sum_{k=1}^{\mathcal{K}}\left \lbrace \textrm{log}_2 ( \mathbf{q}_k^{T}{\mathbf{p}}+\sigma _k^2)+(\widehat{\bb}_k^{(i)})^T {\mathbf{p}} \right \rbrace,
\end{eqnarray}
where
\begin{equation}\label{final156}
\widehat{\bb}_k^{(i)} \triangleq - \frac{\bb_{k}}{ \mathbf{b}_k^{T}{\mathbf{p}}^{(i-1)}+\sigma _k^2} .
\end{equation}
Next, we consider the non-convex constraint in $\textrm{C}_{3}$. Note that $\textrm{C}_{3}$ implies:
\begin{equation}\label{ineq1}
p_{{c}_k} + {{\varepsilon}_k}(1- \tau) {{p}}_{k} \leq \mu_k \tau {\mathbf{x}}^{H}  {{\mathbf{h}}}_{k} {{{\mathbf{h}}}}^{H}_{k}    \mathbf{x} + E_{0,k}.
\end{equation}
This constraint can be deal with via MaMi technique as well. Indeed, we consider the following inequality for a given matrix $\mathbf{T} \succeq \mathbf{0}$ and $\mathbf{x}_0$:
\begin{equation}\label{new1}
{\mathbf{x}}^{H} \mathbf{T} {\mathbf{x}} \geq {\mathbf{x}}^{H}_{0} \mathbf{T} {\mathbf{x}}_{0} +2 \Re \left \lbrace {\mathbf{x}}^{H}_{0} \mathbf{T}  \left( {\mathbf{x}} - {\mathbf{x}}_{0}  \right) \right \rbrace.
\end{equation}
Using the fact that ${{\mathbf{h}}}_{k} {{{\mathbf{h}}}}^{H}_{k}  \succeq \mathbf{0}$, the right-hand side (RHS) of the constraint $\textrm{C}_{3}$ can be substituted at the $i^{th}$ iteration of MaMi by
\begin{align*}\label{ineq200121}
 \mu_k \tau \left( \left( \mathbf{x}^{(i-1)}\right)^{H}  {{\mathbf{h}}}_{k} {{{\mathbf{h}}}}^{H}_{k}    \mathbf{x}^{(i-1)} + 2 \Re \left \lbrace  \left( {{\mathbf{h}}}_{k} {{{\mathbf{h}}}}^{H}_{k}    \mathbf{x}^{(i-1)} \right)^{H}  \left( \mathbf{x} - \mathbf{x}^{(i-1)} \right) \right \rbrace \right) + E_{0,k}.
\end{align*}
Note that the above equation and the minorizer in \eqref{new3} hold for every $k$. Consequently, the problem in \eqref{sum20} can be handled at the $i^{th}$ MaMi iteration by the following problem iteratively:
\begin{eqnarray}\label{final1}
\max_{\mathbf{x},{\mathbf{p}}} && (1-\tau)\sum_{k=1}^{\mathcal{K}} \left \lbrace \textrm{log}_2 ( \mathbf{q}_k^{T}{\mathbf{p}}+\sigma _k^2)+(\widehat{\bb}_k^{(i)})^T {\mathbf{p}} \right \rbrace \\ \nonumber
\mbox{s. t.}\;\;&&
 \textrm{C}_{3}: p_{{c}_k} + {{\varepsilon}_k}(1- \tau) {{p}}_{k} \leq  \mu_k \tau ( \left( \mathbf{x}^{(i-1)}\right)^{H}  {{\mathbf{h}}}_{k} {{{\mathbf{h}}}}^{H}_{k}    \mathbf{x}^{(i-1)}
  \\ \nonumber
&& \hspace{22pt} + 2 \Re \left \lbrace \left({{\mathbf{h}}}_{k} {{{\mathbf{h}}}}^{H}_{k}  \mathbf{x}^{(i-1)} \right)^{H} \left( \mathbf{x} - \mathbf{x}^{(i-1)} \right) \right \rbrace  ) + E_{0,k},\\ \nonumber
&& \textrm{C}_{2}, \textrm{C}_{4}.
\end{eqnarray}

The problem in (\ref{final1}) is a convex optimization and can be solved efficiently by e.g., interior point methods \cite{boyd}.

\subsection{Optimizing $\tau$ for fixed $\mathbf{x}$ and ${\mathbf{p}}$}\label{sumb}
The problem in (\ref{form30}) for fixed energy waveform in phase 1 ($\mathbf{x}$) and transmit powers in phase 2 (${\mathbf{p}}$)
boils down to
\begin{eqnarray}\label{for}
\max_{\tau} && (1 - \tau) \\ \nonumber
\mbox{s. t.}\;\;&&
\textrm{C}_{1}: 0 \leq \tau \leq 1 \\ \nonumber
&& \textrm{C}_{3}: \zeta_{1,k} \leq \tau, \forall k,\\ \nonumber
&& \textrm{C}_{4}: \tau \leq \zeta_{2,k}, \forall k,
\end{eqnarray}
where $\zeta_{1,k} \triangleq\frac{ p_{{c}_k} + {{\varepsilon}_k} {{p}}_{k} - E_{0,k}}{{{\varepsilon}_k} {{p}}_{k} +\mu_k {\mathbf{x}}^{H}  {{\mathbf{h}}}_{k} {{{\mathbf{h}}}}^{H}_{k}    \mathbf{x}   } $, $\zeta_{2,k} \triangleq \frac{E_{{\textrm{max}},{k}} - E_{0,k}}{\mu_k {\mathbf{x}}^{H}  {{\mathbf{h}}}_{k} {{{\mathbf{h}}}}^{H}_{k}    \mathbf{x}}$. Therefore, the closed-form solution for $\tau$ can be expressed as\footnote[1]{For the feasibility of this problem we should have $\zeta_{1,k} \leq 1 , \forall k$, $\zeta_{2,k} \geq 0 , \forall k$, and $\zeta_{1,k} \geq \zeta_{2,k}, \forall k$.}
\begin{equation} \label{for1}
\tau_{\textrm{opt}}= \textrm{max} \lbrace 0, \zeta_{1,k} \rbrace , \forall k.
\end{equation}

\begin{table}[tp]
\footnotesize
\caption{The proposed method for sum throughput maximization in ${\cal K}$-user interference channels} \label{table:method} \centering
\begin{tabular}{p{3.3in}}
\hline \hline
\textbf{Step 0}: Initialize $\tau$ with a random value in $[0,1]$.\\
\textbf{Step 1}: Compute ${\mathbf{x}}^{(\kappa)}$ and ${{\mathbf{p}}}^{(\kappa)}$ by solving the problem in (\ref{sum1}). \\
~~~\textbf{Step 1-1}: Initialize $ {\mathbf{p}} \in \realR^{\cal K}$; set $i=0$.\\
~~~\textbf{Step 1-2}: Solve the convex problem in (\ref{final1}) to obtain the \\~~~most recent version of ${\mathbf{x}}$ and ${\mathbf{p}}$.\\
~~~\textbf{Step 1-3}: Update the parameters in optimization \eqref{final1} and set \\~~~$i=i+1$.\\
~~~\textbf{Step 1-4}: Repeat steps 1-2 and 1-3 till the stop criterion \\~~~is satisfied.\\
\textbf{Step 2}: Compute $\tau^{(\kappa)}$ by solving the problem in (\ref{for}) via the closed-form solution in \eqref{for1}.\\
\textbf{Step 3}: Repeat steps 1 and 2 until a pre-defined stop criterion is satisfied, e.g. $|g^{(\kappa+1)}-g^{(\kappa)}| \leq \xi$ (where $g$ denotes the objective function of the problem (\ref{form30})) for some $\xi>0$.\\
\hline \hline
\end{tabular}
\end{table}

Table~\ref{table:method} summarizes the steps of the proposed method for sum throughput maximization in a ${\cal K}$-user IFC. The devised method consists of outer iterations which are associated with the employed alternating projections approach. At each outer iteration (denoted by superscript $\kappa$), for fixed $\tau$, the convex problem in (\ref{final1}) is solved according to the MaMi iterations i.e., inner iterations (denoted by superscript $i$). Then, for a fixed $\mathbf{x}$ and ${\mathbf{p}}$, the convex optimization (\ref{for}) is handled via the closed-form solution.

\textit{Remark 1 (Convergence of the proposed method)}:
Note that the sequence of objective values of the problem in \eqref{form3} have an ascent property when tackled by the proposed method. More precisely, let $g ({\mathbf{p}}^{(\kappa )}, \mathbf{x}^{(\kappa )} , \tau^{(\kappa )})$ denote the aforementioned objective at the $\kappa^{th}$ iteration. We can write
\begin{equation*}
g \left( {\mathbf{p}}^{(\kappa +1)}, {\mathbf{x}}^{(\kappa +1)}, \tau^{(\kappa +1 )}\right) \geq g \left( {\mathbf{p}}^{(\kappa +1)}, {\mathbf{x}}^{(\kappa +1)}, \tau^{(\kappa  )}\right) \geq g \left( {\mathbf{p}}^{(\kappa +1)}, {\mathbf{x}}^{(\kappa )}, \tau^{(\kappa  )}\right) \geq g \left( {\mathbf{p}}^{(\kappa )}, {\mathbf{x}}^{(\kappa )}, \tau^{(\kappa )}\right) ,
\end{equation*}
where the inequalities above hold due to performing maximization w.r.t. ${\mathbf{p}}$, ${\mathbf{x}}$, and $\tau$, respectively. This property along with the fact that the sum throughput (i.e., $g({\mathbf{p}},{\mathbf{x}} , \tau)$) is bounded above leads to a convergence of the sequence of the objective values. We herein remark on the fact that maximization of $g({\mathbf{p}},{\mathbf{x}} , \tau)$ w.r.t ${\mathbf{p}}$, ${\mathbf{x}}$, and $\tau$ is dealt with via an iterative approach based on the MaMi technique (inner iterations). The convergence of the inner iterations associated with this technique is addressed in the next subsection.
\hfill $\blacksquare$

\subsection{Convergence of the iterations associated with MaMi technique}

We discussed the convergence of the alternating projections method in Remark 1. Therefore, to address the convergence of the proposed method, we need to take into account the inner iteration associated with MaMi technique. Indeed,
consider the original problem $P_0$ in \eqref{mami1} and the problem $P_i$ in \eqref{mami33} resulting from applying MaMi technique at the $i^{th}$ iteration.

First, we show that all iterations of the proposed method is feasible. Let us start with initial point $\widetilde{\bx}^{(0)}$, i.e. $\widetilde{g}(\widetilde{\bx}^{(0)})\leq0$. We should deal with the first iteration by constructing $P_1$. Since according to \eqref{mami2},
$h^{(1)}(\widetilde{\bx}^{(0)})=\widetilde{g}(\widetilde{\bx}^{(0)})\leq0$, $\widetilde{\bx}^{(0)}$ is also feasible for $P_1$ meaning that we can initialize the
iterative procedure by a feasible point to the original problem $P_0$. Let $\widetilde{\bx}^{(i)}$ be a solution to $P_i$. Thus,
$h^{(i)}(\widetilde{\bx}^{(i)})\leq0$. Also, using \eqref{mami2}, we have:
\begin{equation}\label{T1}
\widetilde{g}(\widetilde{\bx}^{(i)})\leq h^{(i)}(\widetilde{\bx}^{(i)})\leq0.
\end{equation}
Noting that $h^{(i+1)}(\widetilde{\bx}^{(i)})=\widetilde{g}(\widetilde{\bx}^{(i)})$ and using \eqref{T1}, we can write the following inequality
\begin{equation}\label{T2}
h^{(i+1)}(\widetilde{\bx}^{(i)})=\widetilde{g}(\widetilde{\bx}^{(i)})\leq h^{(i)}(\widetilde{\bx}^{(i)})\leq0.
\end{equation}
Consequently, the point $\widetilde{\bx}^{(i)}$ is also feasible for the problem $P_{i+1}$ and as a result, the iterations are all feasible.
Next, we show that the sequence of objective values is convergent. Note that we can write
\[ \overbrace {f^{(i+1)}(\widetilde{\bx}) =  q^{(i+1)}(\widetilde{\bx})}^{\text{minorizer property}} \geq  \overbrace {q^{(i)}(\widetilde{\bx}) \geq f^{(i)}(\widetilde{\bx})}^{\text{minorizer property}}. \]
\vspace{-47pt}
\[ \phantom{h^{(i+1)}(\widetilde{\bx}) =} \underbrace{\phantom{ p^{(i+1)}(\widetilde{\bx}) \geq  p^{(i)}(\widetilde{\bx}) }}_{\text{ maximization step}}  \phantom{\geq h^{(i)}(\widetilde{\bx}) } \addtag \label{T3} \]
The inequality above holds due to maximization step and note that the point $\widetilde{\bx}^{(i)}$ is a feasible point for $P_{i+1}$ (see \eqref{T2}), but in general, the optimal solution of $P_{i+1}$, i.e. $\widetilde{\bx}^{(i+1)}$ has larger objective function value when compared to this feasible point.
Using \eqref{T3} and noting that the objective function $f(.)$ is upper-bounded, we conclude that the sequence of objective values converges.

\section{Achieving Fairness in the network: maxmin throughput Optimization} \label{maxmin}
In Section III, the sum throughput of the network was maximized by judicious selection of the allocated WET-WIT time slot $\tau$, the energy waveform in phase 1 ($\mathbf{x}$), and the transmit powers ${\mathbf{p}}$ in phase 2. However, there might be user pairs suffering from a low throughput while some others have a high throughput for their communications. To improve the throughput of the pairs with low throughput and hence achieve a fairness in the network, in this section, we aim to improve the throughput of the worst user pair (i.e., the pair associated with $\displaystyle \min_{k}~R_k({\mathbf{p}},\tau)$) by optimizing the energy waveform in first phase and the transmit powers in second phase as well as the allocated time slot $\tau$. Therefore, we cast the following max-min throughput optimization problem
\begin{eqnarray}\label{maxmin1}
\max_{\mathbf{x},{\mathbf{p}},\tau} & \displaystyle \min_{1 \leq k \leq {\cal K}} & (1 - \tau) \textrm{log}_2 \left(1+\gamma_k ( {\mathbf{p}} )\right)\\ \nonumber
\mbox{s. t.}\;\;&&
\textrm{C}_{1}, \textrm{C}_{2}, \textrm{C}_{3}, \textrm{C}_{4},
\end{eqnarray}
where $\gamma_k ({\mathbf{p}})$ is defined in \eqref{tr}.

Note that the problem in \eqref{maxmin1} is non-convex due to the coupled design variables in the objective function and the constraint set. To tackle this design problem, similar to the sum throughput maximization, we employ an alternating projections approach with the partitioning similar to that in Section \ref{sum}. Again, the resulted problems are non-convex and we aim to devise efficient algorithms to solve them suboptimally.

\subsection{Optimizing $\mathbf{x}$ and ${\mathbf{p}}$ in max-min problem for fixed $\tau$} \label{maxa}
For fixed $\tau$, the problem \eqref{maxmin1} boils down to the following problem:
\begin{eqnarray}\label{maxmin21}
\max_{\mathbf{x},{\mathbf{p}}} & \displaystyle \min_{1 \leq k \leq {\cal K}} & (1 - \tau) \textrm{log}_2 \left( \frac{ \mathbf{q}_k^{T}{\mathbf{p}}+\sigma _k^2}{ \mathbf{b}_k^{T}{\mathbf{p}} +\sigma _k^2}\right)\\ \nonumber
\mbox{s. t.}\;\;&& \textrm{C}_{2}, \textrm{C}_{3}, \textrm{C}_{4},
\end{eqnarray}
where $\mathbf{q}_{k}$ and $\mathbf{b}_{k}$ are given in Section \ref{suma}.

The problem above can be equivalently rewritten with an auxiliary variable $\alpha$ as
\begin{eqnarray}\label{maxmin321}
\max_{\mathbf{x},{\mathbf{p}},\alpha} \;\;&& \alpha \\ \nonumber \mbox{s. t.}\;\;&&  \textrm{C}_{2}, \textrm{C}_{3}, \textrm{C}_{4}, \\ \nonumber \;\;&&  \textrm{C}_{5}: (1-\tau) \left\lbrace \textrm{log}_2 ( \mathbf{q}_k^{T}{\mathbf{p}}+\sigma _k^2) - \textrm{log}_2 ( \mathbf{b}_k^{T}{\mathbf{p}}+\sigma _k^2) \right\rbrace \geq \alpha , \forall k.
\end{eqnarray}
Let $\widetilde{f}_{1,k} ({\mathbf{p}}) \triangleq \textrm{log}_2 ( \mathbf{q}_k^{T}{\mathbf{p}}+\sigma _k^2) $ and $\widetilde{f}_{2,k} ({\mathbf{p}}) \triangleq - \textrm{log}_2 ( \mathbf{b}_k^{T}{\mathbf{p}}+\sigma _k^2)$. Next, observe that $\widetilde{f}_{1,k} ({\mathbf{p}}) $ and $\widetilde{f}_{2,k} ({\mathbf{p}}) $ are concave and convex functions w.r.t. ${\mathbf{p}}$, respectively for all $k$. Consequently, the constraint $\textrm{C}_{5}$ in the \eqref{maxmin321}, i.e., $\widetilde{f}_{1,k}({\mathbf{p}})+\widetilde{f}_{2,k}({\mathbf{p}}) \geq \alpha, \forall~k$ is non-convex. Note that the constraint $\textrm{C}_{3}$ is also non-convex but can be dealt with similarly as in Section \ref{suma}, denoted by $\widehat{\textrm{C}}^{(i)}_{3}$ in the sequel. To deal with the non-convexity of $\textrm{C}_{5}$, we apply MaMi technique again, i.e.,.
by replacing the convex function $\widetilde{f}_{2,k}$ with a concave (or linear) lower bound iteratively (for each $k$) such that the resulted constraint set becomes a convex set at each iteration (see  (\ref{mami1})- (\ref{mami33})). 
Due to the fact that the function $\widetilde{f}_{2,k}{({\mathbf{p}})}$ is convex w.r.t. ${\mathbf{p}}$, it can be minorized using its supporting hyperplane. This minorizer can be obtained by employing the aforementioned inequality in \eqref{ineq} for $x \triangleq \mathbf{b}_k^{T}{\mathbf{p}} +\sigma _k^2$.
This leads to
\begin{equation*}
- \textrm{log}_2 ( \mathbf{b}_k^{T}{\mathbf{p}}+\sigma _k^2) \geq - \textrm{log}_2 ( \mathbf{b}_k^{T}{{\mathbf{p}}}_0+\sigma _k^2) + \widehat{\bb}_k^T({\mathbf{p}}-{{\mathbf{p}}}_0),
\end{equation*}
where $\widehat{\bb}_k$ has been defined in \eqref{final156}. Substituting the above minorizer in lieu of $\widetilde{f}_{2,k}({\mathbf{p}})$ in the constraint set of the problem in \eqref{maxmin321} (for each $k$) and employing $\widehat{\textrm{C}}^{(i)}_{3}$ lead to the below optimization at the $i^{th}$ iteration of the MaMi technique,
\begin{eqnarray}\label{maxmin32}
\max_{\mathbf{x},{\mathbf{p}},\alpha} \;\;&& \alpha \\ \nonumber \mbox{s. t.}\;\;&& \hspace{-8pt} \textrm{C}_{2}, \widehat{\textrm{C}}^{(i)}_{3}, \textrm{C}_{4},
\\ \nonumber \;\;&& \hspace{-8pt} \textrm{C}_{5}:(1- \tau) \left\lbrace \textrm{log}_2 ( \mathbf{q}_k^{T} {\mathbf{p}}+\sigma _k^2) - \textrm{log}_2 ( \mathbf{b}_k^{T}{{\mathbf{p}}}^{(i-1)}+\sigma _k^2) +
(\widehat{\bb}_k^{(i)})^{T}({\mathbf{p}}-{{\mathbf{p}}}^{(i-1)}) \right\rbrace \geq \alpha , \forall k.
\end{eqnarray}
Note that the problem in (\ref{maxmin32}) is now a convex optimization and can be solved efficiently via e.g., interior point methods.

\subsection{Optimizing $\tau$ in max-min problem for fixed $\mathbf{x}$ and ${\mathbf{p}}$}\label{maxb}
For fixed $\mathbf{x}$ and ${\mathbf{p}}$, and using the fact that
\begin{equation*}
\displaystyle \min_{1\leq k \leq \mathcal{K}}  \hspace{5pt} (1 - \tau) \textrm{log}_2 \left( \frac{ \mathbf{q}_k^{T}{\mathbf{p}}+\sigma _k^2}{ \mathbf{b}_k^{T}{\mathbf{p}} +\sigma _k^2}\right) = 1 - \tau,
\end{equation*}
  the maxmin problem in (\ref{maxmin1}) boils down to the problem in \eqref{for} in Section \ref{sumb} with the  closed-form solution in \eqref{for1}.
\begin{table}[tp]
\footnotesize
\caption{The proposed method for max-min throughput optimization in ${\cal K}$-user interference channel} \label{table:method2} \centering
\begin{tabular}{p{3.3in}}
\hline \hline
\textbf{Step 0}: Initialize $\tau$ with random values in $[0,1]$. \\
\textbf{Step 1}: Compute ${\alpha}^{(\kappa)}$, ${\mathbf{x}}^{(\kappa)}$, and ${{\mathbf{p}}}^{(\kappa)}$ by solving the problem in (\ref{maxmin321}):\\
~~~\textbf{Step 1-1}: Initialize $ {\mathbf{p}} \in \realR^{\cal K}$ and; set $i=0$.\\
~~~\textbf{Step 1-2}: Solve the problem in (\ref{maxmin32}) to obtain the most \\~~~recent version of ${\alpha}$, ${\mathbf{x}}$, and ${\mathbf{p}}$.\\
~~~\textbf{Step 1-3}: Update the parameters in optimization \eqref{maxmin32} and set \\~~~$i=i+1$.\\
~~~\textbf{Step 1-4}: Repeat steps 1-2 and 1-3 till the stop criterion \\~~~is satisfied.\\
\textbf{Step 2}: Compute $\tau^{(\kappa)}$ via the closed-form solution in \eqref{for1}:\\
\textbf{Step 3}: Repeat steps 1 and 2 until a pre-defined stop criterion is satisfied, e.g. $|g_w^{(\kappa+1)}-g_w^{(\kappa)}| \leq \xi_w$ (where $g_w$ denotes the objective function of the problem (\ref{maxmin1})) for some $\xi_w>0$.\\
\hline \hline
\end{tabular}
\end{table}

The steps of the devised algorithm for max-min throughput optimization are summarized in Table~\ref{table:method2}. Similar to the sum throughput maximization, the proposed method consists of outer iterations (denoted by superscript $\kappa$) which are associated with the employed alternating projections approach. At each outer iteration, for fixed $\tau$, the convex problem in (\ref{maxmin32}) is solved according to the MaMi iterations i.e., inner iterations (denoted by superscript $i$). Then, for a fixed $\mathbf{x}$ and ${\mathbf{p}}$, the optimal value of $\tau$ is obtained via the closed-form solution in \eqref{for1}.
\section{further notes and discussions}
\subsection{Non-linearity of EH circuit} \label{nonn}
\begin{figure}
\centering
\includegraphics[scale=.6]{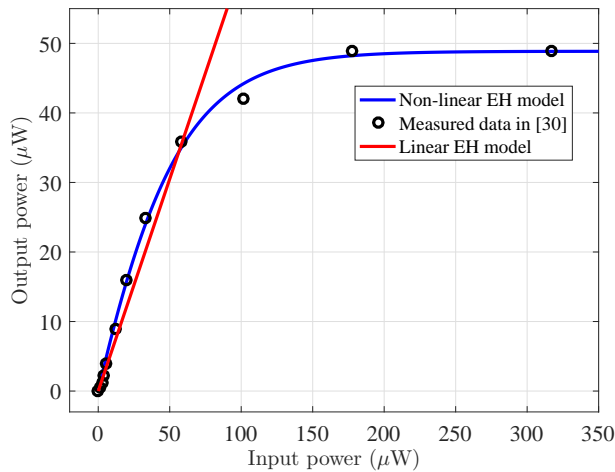}
\caption{A comparison between the linear and non-linear EH models with practical data from Fig. 7 of \cite{ungan2009rf}.}
\label{P22}
\centering
\end{figure}
In the previous sections, we considered the linear model for EH circuits given in (2). However,
in practice, there is a non-linear characteristic between input and output powers of the EH
circuits, especially in the saturation region. For example, as shown in Fig. \ref{P22}, there is a considerable difference between linear and non-linear curves in the saturation region, whereas the aforementioned models have minor differences in the linear region.
Herein, we consider the non-linear EH model for ERs in the first phase. In
this case, a modified version of \eqref{Q} for the harvested energy by the $k^{th}$ ER for non-linear model
can be expressed as \cite{brobust}:
\begin{equation}\label{hh}
E^{nl}_k (\mathbf{x})=\tau  \frac{\frac{N_k}{1+\textrm{exp} \left(-\widetilde{a}_k \left(  p^l_k (\mathbf{x})   -\widetilde{b}_k \right)  \right)}-N_k \Omega_k}{1- \Omega_k} , \hspace{5pt} \Omega_k=\frac{1}{1+\textrm{exp} \left(\widetilde{a}_k \widetilde{b}_k \right)} ,  \hspace{8pt} 1 \leq k \leq \mathcal{K},
\end{equation}
where, $p^l_k (\mathbf{x}) =  {\mathbf{x}}^{H}  {{\mathbf{h}}}_{k} {{{\mathbf{h}}}}^{H}_{k}    \mathbf{x}$, $N_k$ is the maximum power that each ER can harvest and $\widetilde{a}_k$ as well as $\widetilde{b}_k$ are the factors for non-idealities of the EH circuit. As an example, the values of the parameters $N_k=48.86 ~\mu W$, $\widetilde{a}_k=26515.46$ and $\widetilde{b}_k=-29.81 \times 10^{-6}$ of the model in \eqref{hh} can be obtained via curve fitting tools with the practical data from \cite{ungan2009rf} with the high R-squared value of $R^2=0.9981$\footnote[1]{Note that the curve fitting can also be performed in log-log scale to obtain more accurate results in low power regimes (see \cite{clerckx2019fundamentals} for details).}.
Therefore, the constraints $\textrm{C}_{3}$ and $\textrm{C}_{4}$ of design problems \eqref{form3} and \eqref{maxmin1} can be reformulated as
\begin{equation*}
\textrm{C}^{nl}_{3}: p_{{c}_k} + {{\varepsilon}_k}(1- \tau) {{p}}_{k} \leq E^{nl}_k (\mathbf{x}) + E_{0,k}, \forall k,
\end{equation*}
\begin{equation*}
 \textrm{C}^{nl}_{4}: E^{nl}_k (\mathbf{x})+ E_{0,k} \leq E_{{\textrm{max}},{k}}, \forall k.
\end{equation*}
Note that $E^{nl}_k (\mathbf{x})$ is a non-decreasing concave function w.r.t. $p^{l}_k (\mathbf{x})$ (for typical values of $N_k$, $\widetilde{a}_k$ and $\widetilde{b}_k$) and $p^{l}_k (\mathbf{x})$ is a convex function w.r.t. $\mathbf{x}$; therefore, $E^{nl}_k (\mathbf{x})$ is neither convex nor concave w.r.t. $\mathbf{x}$. Consequently, both constraints in $\textrm{C}^{nl}_{3}$ and $\textrm{C}^{nl}_{4}$ represent non-convex sets. To deal with such non-convex sets, we observe that $E^{nl}_k (\mathbf{x})$ can be rewritten as a sum of a convex and a concave function for sufficiently large\footnote[2]{See Appendix \ref{app1} for a selection of $\beta$.} $\beta$ as stated below \cite[Theorem 1]{yuille2002concave}
\begin{equation} \label{jj}
 E^{nl}_k (\mathbf{x})= \underbrace{ E^{nl}_k (\mathbf{x}) + \frac{1}{2} \beta {\mathbf{x}}^{H} \mathbf{x}}_{\text{convex}}
 \underbrace{- \frac{1}{2} \beta {\mathbf{x}}^{H} \mathbf{x}}_{\text{concave}}, ~\forall k.
\end{equation}
Now, for the non-convex constraint $\textrm{C}^{nl}_{3}$, we can keep the concave part in \eqref{jj} and minorize the convex part to obtain a convex constraint. For $\textrm{C}^{nl}_{4}$, we may keep the convex part in \eqref{jj} and majorize the concave part using the MaMi technique. The aforementioned convex/concave parts can be minorized/majorized using the definition of the convexity, as stated below
\begin{equation*}\label{in}
\textrm{C}^{nl}_{3}:  p_{{c}_k} + {{\varepsilon}_k}(1- \tau) {{p}}_{k} \leq  E^{nl}_k ({\mathbf{x}}^{(i-1)}) + \frac{1}{2} \beta ({{\mathbf{x}}^{(i-1)}})^{H} {\mathbf{x}}^{(i-1)} + \Re \left \lbrace {\mathbf{u}}^{(i)}_{k}   \left( \mathbf{x} - {\mathbf{x}}^{(i-1)} \right) \right \rbrace -\frac{1}{2} \beta {\mathbf{x}}^{H} \mathbf{x} + E_{0,k},
\end{equation*}
\begin{equation*}\label{ine}
\textrm{C}^{nl}_{4}:   E^{nl}_k(\mathbf{x}) +\frac{1}{2} \beta {\mathbf{x}}^{H} \mathbf{x} -\frac{1}{2} \beta \Big( ({{\mathbf{x}}^{(i-1)}})^{H} {\mathbf{x}}^{(i-1)} +2  \Re \left \lbrace ({{\mathbf{x}}^{(i-1)}})^{H}  \left( \mathbf{x} - {{\mathbf{x}}^{(i-1)}} \right) \right \rbrace \Big)+ E_{0,k}  \leq E_{{\textrm{max}},{k}},
\end{equation*}
where, ${\mathbf{u}}^{(i)}_{k}$ can be expressed as
\begin{equation*}
{\mathbf{u}}^{(i)}_{k} = \frac{2\tau N_k \widetilde{a}_k \textrm{exp} \left(-\widetilde{a}_k \left(  ({{\mathbf{x}}^{(i-1)}})^{H} {{\mathbf{h}}}_{k} {{{\mathbf{h}}}}^{H}_{k}   {\mathbf{x}}^{(i-1)}   -\widetilde{b}_k \right)  \right)}{(1-\Omega_k) \left(1+\textrm{exp} \left(-\widetilde{a}_k \left( ({{\mathbf{x}}^{(i-1)}})^{H} {{\mathbf{h}}}_{k} {{{\mathbf{h}}}}^{H}_{k}   {\mathbf{x}}^{(i-1)} -\widetilde{b}_k \right)  \right) \right) ^{2}} ~ \left({{\mathbf{x}}^{(i-1)}}\right)^{H} {{\mathbf{h}}}_{k} {{{\mathbf{h}}}}^{H}_{k}  +\beta \left({{\mathbf{x}}^{(i-1)}} \right)^{H}.
\end{equation*}
\subsection{Effect of imperfect CSI}
In practice, the perfect CSI is not available and thus there might be uncertainties w.r.t. channel coefficients $h_{j,k}$ and $g_{j,k}$.  Therefore, in this subsection, we consider imperfect CSI for optimizing the design parameters.
The imperfect CSI between ET-IT pairs using the linear minimum mean squared error (LMMSE) estimator can be modeled as \cite{vaezy2018efficient}
\begin{equation}
h_{k,j} = \widehat{h}_{k,j} + \Delta h_{k,j},
\label{ratek1}
\end{equation}
\begin{equation*}
g_{k,j} = \widehat{g}_{k,j} + \Delta g_{k,j},
\label{ratek100}
\end{equation*}
where $\widehat{h}_{k,j}$ and $\widehat{g}_{k,j}$ are the estimates of channel coefficients ${h}_{k,j}$ and ${g}_{k,j}$, respectively, while $\Delta h_{k,j}$ and $\Delta g_{k,j}$ are the channel estimation errors. It is assumed that $\Delta h_{k,j}$ as well as $\Delta g_{k,j}$ are independent zero-mean complex Gaussian random variables with variance $\sigma^{2}_{h,\Delta}$ and $\sigma^{2}_{g,\Delta}$, and also are uncorrelated with ${\widehat{h}}_{k,j}$ and ${\widehat{g}}_{k,j}$, respectively. The relationships between the variances of the channel coefficients $\sigma^{2}_{h}$ and $\sigma^{2}_{g}$, the variances of the estimated channel coefficients $\sigma^{2}_{\widehat{h}}$ and $\sigma^{2}_{\widehat{g}}$, and the variances of the channel estimation errors can be expressed as
\begin{equation}
\sigma^{2}_{h,\Delta} =(1-{{\rho}_{h}}^{2}) \sigma^{2}_{h}, \hspace{10pt}\sigma^{2}_{\widehat{h}}= {{\rho}_{h}}^{2} \sigma^{2}_{h},
\label{ratek1001}
\end{equation}
\begin{equation*}
\sigma^{2}_{g,\Delta} =(1-{{\rho}_{g}}^{2}) \sigma^{2}_{g}, \hspace{10pt}\sigma^{2}_{\widehat{g}}= {{\rho}_{g}}^{2} \sigma^{2}_{g},
\label{ratek1002}
\end{equation*}
where the parameters $ {\rho}_{g}, {\rho}_{h}  \in [0,1]$ indicate the estimation accuracy.
According to \eqref{ratek1} and \eqref{ratek1001}, the amount of harvested energy in \emph{average sense} can be computed using the fact that the channel estimate $\widehat{\mathbf{h}}_{k}$ is available at the begining of each block of duration $T$ as \cite[eq.~(23)]{zeng2015optimized}
\begin{align}
E^{im}_k &=\mu_k \tau {\mathbb{E}}_{\Delta {{\mathbf{h}}}_{k}} \lbrace y_k y^{H}_{k} \vert \widehat{{\mathbf{h}}}_{k} \rbrace =\mu_k \tau {\mathbb{E}}_{\Delta {{\mathbf{h}}}_{k}} \lbrace \mathbf{h}^{H}_{k} \mathbf{x} {\mathbf{x}}^{H} {\mathbf{h}}_{k} \vert \widehat{{\mathbf{h}}}_{k} \rbrace =\mu_k \tau \textrm{tr} \left\lbrace {\mathbb{E}}_{\Delta {{\mathbf{h}}}_{k}} \lbrace \mathbf{h}^{H}_{k} \mathbf{x} {\mathbf{x}}^{H} {\mathbf{h}}_{k} \vert \widehat{{\mathbf{h}}}_{k} \rbrace \right\rbrace \\ \nonumber & =\mu_k \tau \textrm{tr} \left\lbrace {\mathbb{E}}_{\Delta {{\mathbf{h}}}_{k}} \left \lbrace {\mathbf{h}}_{k} \mathbf{h}^{H}_{k}\vert \widehat{{\mathbf{h}}}_{k} \right \rbrace  \mathbf{x} {\mathbf{x}}^{H}  \right\rbrace =\mu_k \tau {\mathbf{x}}^{H} {\mathbb{E}}_{\Delta {{\mathbf{h}}}_{k}} \left \lbrace ( \widehat{{\mathbf{h}}}_{k} +\Delta {{\mathbf{h}}}_{k} ) ({\widehat{{\mathbf{h}}}}^{H}_{k} + \Delta {{\mathbf{h}}}^{H}_{k} ) \vert \widehat{{\mathbf{h}}}_{k} \right \rbrace  \mathbf{x}  \\ \nonumber & = \mu_k \tau {\mathbf{x}}^{H}  \left( \widehat{{\mathbf{h}}}_{k} {\widehat{{\mathbf{h}}}}^{H}_{k} + {\mathbb{E}}_{\Delta {{\mathbf{h}}}_{k}} \left \lbrace \Delta {{\mathbf{h}}}_{k}   \Delta {{\mathbf{h}}}^{H}_{k} \right \rbrace \right)   \mathbf{x}= \mu_k \tau {\mathbf{x}}^{H}  \left( \widehat{{\mathbf{h}}}_{k} {\widehat{{\mathbf{h}}}}^{H}_{k} + \sigma^{2}_{h,\Delta} \mathbf{I}_{\mathcal{K}} \right)   \mathbf{x},
\end{align}
where $\widehat{\mathbf{h}}_{k}\triangleq{[\widehat{h}^{*}_{k,1} ,\widehat{h}^{*}_{k,2}, ... ,\widehat{h}^{*}_{k,\mathcal{K}}]}^{T}$ and $\Delta\mathbf{h}_{k}\triangleq{[\Delta{h}^{*}_{k,1} ,\Delta{h}^{*}_{k,2}, ... ,\Delta{h}^{*}_{k,\mathcal{K}}]}^{T}$.
The SINR of the $k^{th}$ ET-IT pair is then given by 
\begin{align}
\gamma^{im}_k ({\mathbf{p}}) &=\frac{ \vert \widehat{g}_{k,k} \vert ^2 {{p}}_k}{ \sum_{j=1, k\neq j}^{\mathcal{K}} \vert \widehat{g}_{j,k} \vert ^2 {{p}}_{j}+\sigma^{2}_{g,\Delta} {\sum_{j=1}^{\mathcal{K}} {{p}}_{j}}+ \sigma _k^2}.
\end{align}
Therefore, the throughput of the $k^{th}$ ET-IT pair and also constraints $\textrm{C}_{3}$ and $\textrm{C}_{4}$ of design problems \eqref{form3} and \eqref{maxmin1} can be reformulated as
\begin{equation}
R^{im}_k ({\mathbf{p}}, \tau) = (1 - \tau) \textrm{log}_2 \left(1 + \gamma^{im}_k ({\mathbf{p}}) \right)= (1 - \tau) \textrm{log}_2 \left(1 + \frac{ \left( {\mathbf{a}}^{im}_k  \right)^{T}{\mathbf{p}}}{\left( {\mathbf{b}}^{im}_k  \right)^{T}{\mathbf{p}}+ \sigma _k^2} \right),
\end{equation}
\begin{equation*}
\textrm{C}^{im}_{3}: p_{{c}_k} + {{\varepsilon}_k}(1- \tau) {{p}}_{k} \leq E^{im}_k (\mathbf{x}) + E_{0,k}, \forall k,
\end{equation*}
\begin{equation*}
 \textrm{C}^{im}_{4}: E^{im}_k (\mathbf{x})+ E_{0,k} \leq E_{{\textrm{max}},{k}}, \forall k.
\end{equation*}
respectively, where ${\mathbf{a}}^{im}_k\triangleq {a}^{im}_{k} \mathbf{e_k}$ and ${\mathbf{b}}^{im}_k\triangleq [{b}^{im}_{k,1}, {b}^{im}_{k,2}, ... , {b}^{im}_{k,\mathcal{K}}]^T$ with
\begin{equation*}
a^{im}_{k} \triangleq \vert \widehat{g}_{k,k} \vert ^2 , \hspace{.5cm}
b^{im}_{k,j} \triangleq
\begin{cases}
\vert \widehat{g}_{j,k} \vert ^2 +\sigma^{2}_{g,\Delta} , \hspace{.2cm} k\neq j,
\\
\sigma^{2}_{g,\Delta} ,\hspace{.5cm} k= j.
\end{cases}
\end{equation*}
Note that the optimization problems in \eqref{form30} and \eqref{maxmin1} can be recast and solved in case of imperfect CSI,
similar to the procedures in Table~\ref{table:method} and Table~\ref{table:method2} with the above expressions.

\section{Numerical Examples} \label{simu}
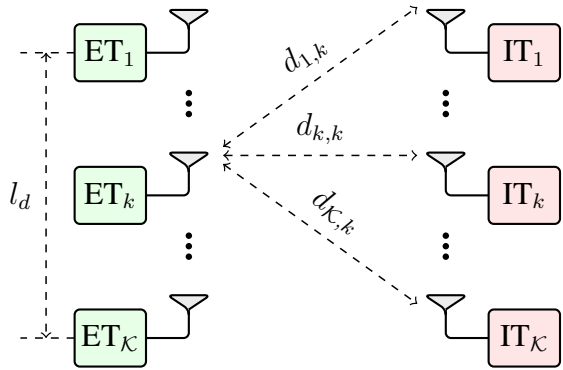
\begin{figure}
\centering
\begin{tikzpicture}[even odd rule,rounded corners=2pt,x=12pt,y=12pt,scale=.9]
\draw[thick,fill=green!10] (-15,5) rectangle ++(2.5,2) node[midway]{$\textrm{ET}_1$};
\draw[thick] (-12.5,6)--++(1.5,0)--+(0,1);
\draw[thick,fill=gray!15] (-11,7)--++(.75,0.5)--++(-1.5,0)--++(.75,-.5)--++(0,-.1);
\draw (-11,4.5) node {\huge$\vdots$};
\draw[thick,fill=green!10] (-15,0) rectangle ++(2.5,2) node[midway]{$\textrm{ET}_k$};
\draw[thick] (-12.5,1)--++(1.5,0)--+(0,1);
\draw[thick,fill=gray!15] (-11,2)--++(.75,0.5)--++(-1.5,0)--++(.75,-.5)--++(0,-.1);
\draw (-11,-.5) node {\huge$\vdots$};
\draw[<->,dashed,line width=.2mm,black!100] (-9.8,2.7)--++(6.8,4.9) node[pos=0.5,above,sloped]{$d_{1,k}$};
\draw[<->,dashed,line width=.2mm,black!100] (-9.8,2.1)--++(6.8,-4.9)node[pos=0.5,above,sloped]{$d_{\mathcal{K},k}$};
\draw[<->,dashed,line width=.2mm,black!100] (-9.8,2.4)--++(6.8,0)node[pos=0.5,above,sloped]{$d_{k,k}$};
\draw[thick,fill=green!10] (-15,-5) rectangle ++(2.5,2) node[midway]{$\textrm{ET}_\mathcal{K}$};
\draw[thick] (-12.5,-4)--++(1.5,0)--+(0,1);
\draw[thick,fill=gray!15] (-11,-3)--++(.75,0.5)--++(-1.5,0)--++(.75,-.5)--++(0,-.1);
\draw[-,dashed,line width=.2mm,black!100](-15,-4)--++(-1.9,0);
\draw[-,dashed,line width=.2mm,black!100](-15,6)--++(-1.9,0);
\draw[<->,dashed,line width=.2mm,black!100](-16,6)--++(0,-10);
\node [] at (-16.9,1){$l_{d}$};
\draw[thick,fill=red!10] (-.5,5) rectangle ++(2.5,2) node[midway]{ $\textrm{IT}_1$};
\draw[thick] (-.5,6)--++(-1.5,0)--+(0,1);
\draw[thick,fill=gray!15] (-2,7)--++(.75,0.5)--++(-1.5,0)--++(.75,-.5)--++(0,-.1);
\draw (-2,4.5) node {\huge$\vdots$};
\draw[thick,fill=red!10] (-.5,0) rectangle ++(2.5,2) node[midway]{ $\textrm{IT}_k$};
\draw[thick] (-.5,1)--++(-1.5,0)--+(0,1);
\draw[thick,fill=gray!15] (-2,2)--++(.75,0.5)--++(-1.5,0)--++(.75,-.5)--++(0,-.1);
\draw (-2,-.5) node {\huge$\vdots$};
\draw[thick,fill=red!10] (-.5,-5) rectangle ++(2.5,2) node[midway]{ $\textrm{IT}_\mathcal{K}$};
\draw[thick] (-.5,-4)--++(-1.5,0)--+(0,1);
\draw[thick,fill=gray!15] (-2,-3)--++(.75,0.5)--++(-1.5,0)--++(.75,-.5)--++(0,-.1);
\end{tikzpicture}
\caption{A $\mathcal{K}$-link symmetric IFC.}
\label{finalom}
\centering
\end{figure}

In this section, the performance of the proposed method is evaluated via Monte-Carlo simulations.
Without loss of generality, we assume that the channel reciprocity holds for the phase 1 and phase 2, i.e., $h_{i,j} = g_{i,j} , i,j = 1, ... , \mathcal{K}$ \cite{Throughput}. Channel coefficients $\lbrace h_{i,j} \rbrace$ are modeled according to Rician fading in which the complex channel between $j^{th}$ ET and $i^{th}$ IT is given by \cite{coll}
\begin{equation}
h_{i,j} = \Big[ \sqrt{\frac{M}{M+1}}h^{\textrm{LoS}} + \sqrt{\frac{1}{M+1}} h^{\textrm{NLoS}}_{i,j} \Big] \sqrt{c_0 (\frac{d_{i,j}}{d_0})^{-v}},\hspace{8pt} \forall i,j,
\end{equation}
where $ h^{\textrm{LoS}} $ is the line-of-sight (LoS) deterministic component with $ \vert h^{\textrm{LoS}} \vert ^2 =1$, $ h^{\textrm{NLoS}}_{i,j} $ is a circularly symmetric complex Gaussian random variable with zero mean and unit variance representing non-LoS Rayleigh fading component, $ M $ denotes the Rician factor, $ c_0 $ is a constant attenuation due to the path-loss at a reference distance $ d_0$, $ v$ is a path loss exponent and $ d_{i,j} $ is the distance between the $j^{th}$ ET and the $i^{th}$ IT. Throughout the simulation, we consider $M=3, c_0=-20$ $\textrm{dB}, d_0=1$ meter (m), $v=3$, and $\sigma _k^2 = -70$ $\textrm{dBm} ,\forall k $\cite{coll}. We consider both symmetric and asymmetric scenarios. In the symmetric scenario (shown in Fig. \ref{finalom}), we limit all pairs in $l_d=50$ (m) such that the distance between any pair and its adjacent pair is equal to $\frac{l_d}{\mathcal{K}-1}$ unless otherwise explicitly stated. Also, we set $p_{\textrm{max},k} = 32$ dBm, $\forall k $, $p_{{c}_k}=-23$ dBm, $\forall k $, ${\varepsilon}_k=1$, $\forall k $, the estimation quality $ {\rho}_{g}={\rho}_{h} =\rho=0.9$, and $d_{k,k} =10$ (m), $\forall k$. We further assume the number of pairs $\mathcal{K}=5$, unless otherwise explicitly stated. In addition, we set $E_{0,{k}}= 0 $ J, $\forall k $, and $E_{{\textrm{max}},{k}}= 50$ $\mu$J, $\forall k $ \cite{lee2016resource}. We also investigate an asymmetric scenario for evaluating the performance of the proposed system under more general condition as shown in Fig. \ref{farIT} (in Section \ref{asym}). The convex optimization problems are solved using CVX package \cite{cvx}.

\subsection{Convergence of proposed algorithms}
\begin{figure}
\hfill
\subfigure[]{\includegraphics[width=8cm]{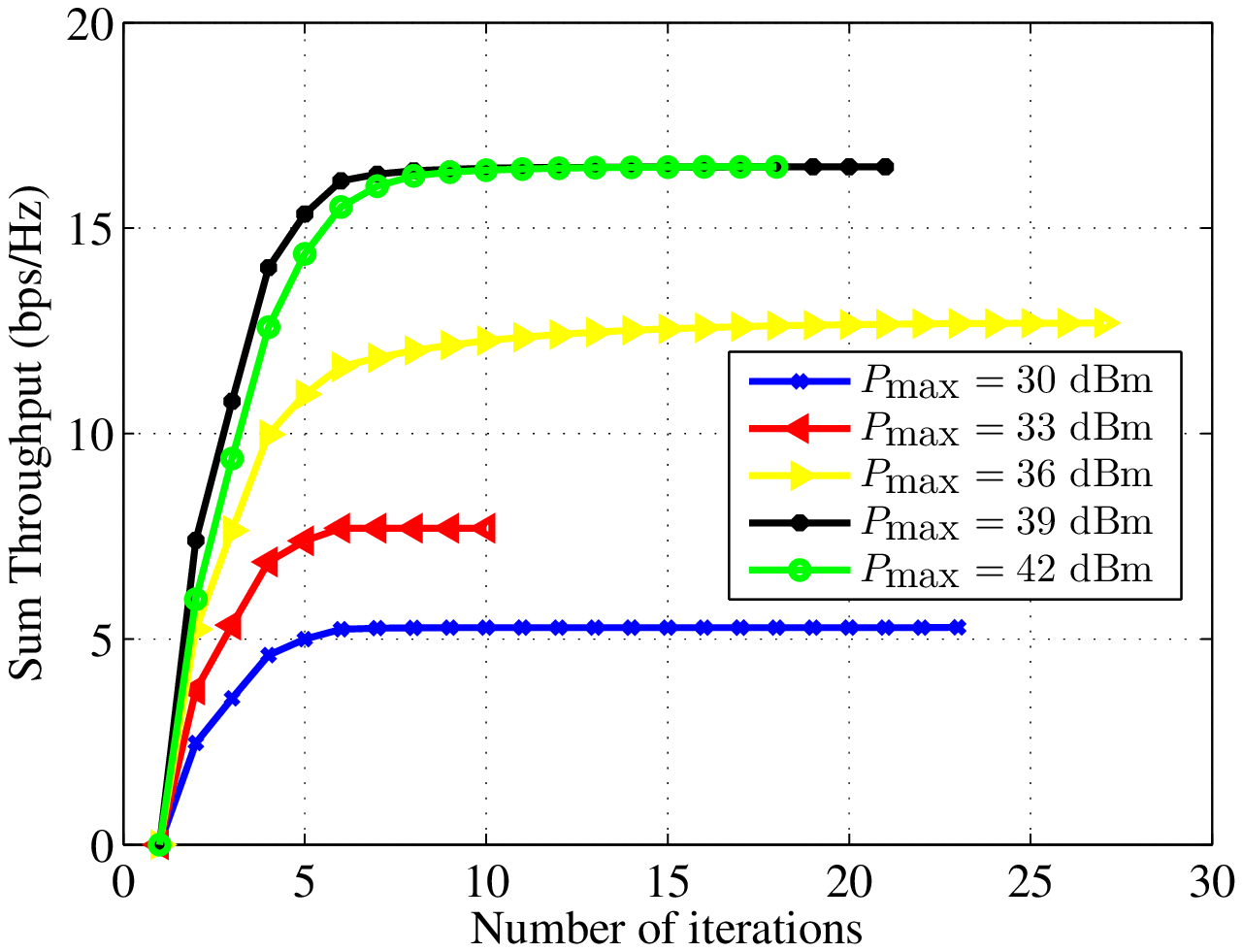}}
\hfill
\subfigure[]{\includegraphics[width=8cm]{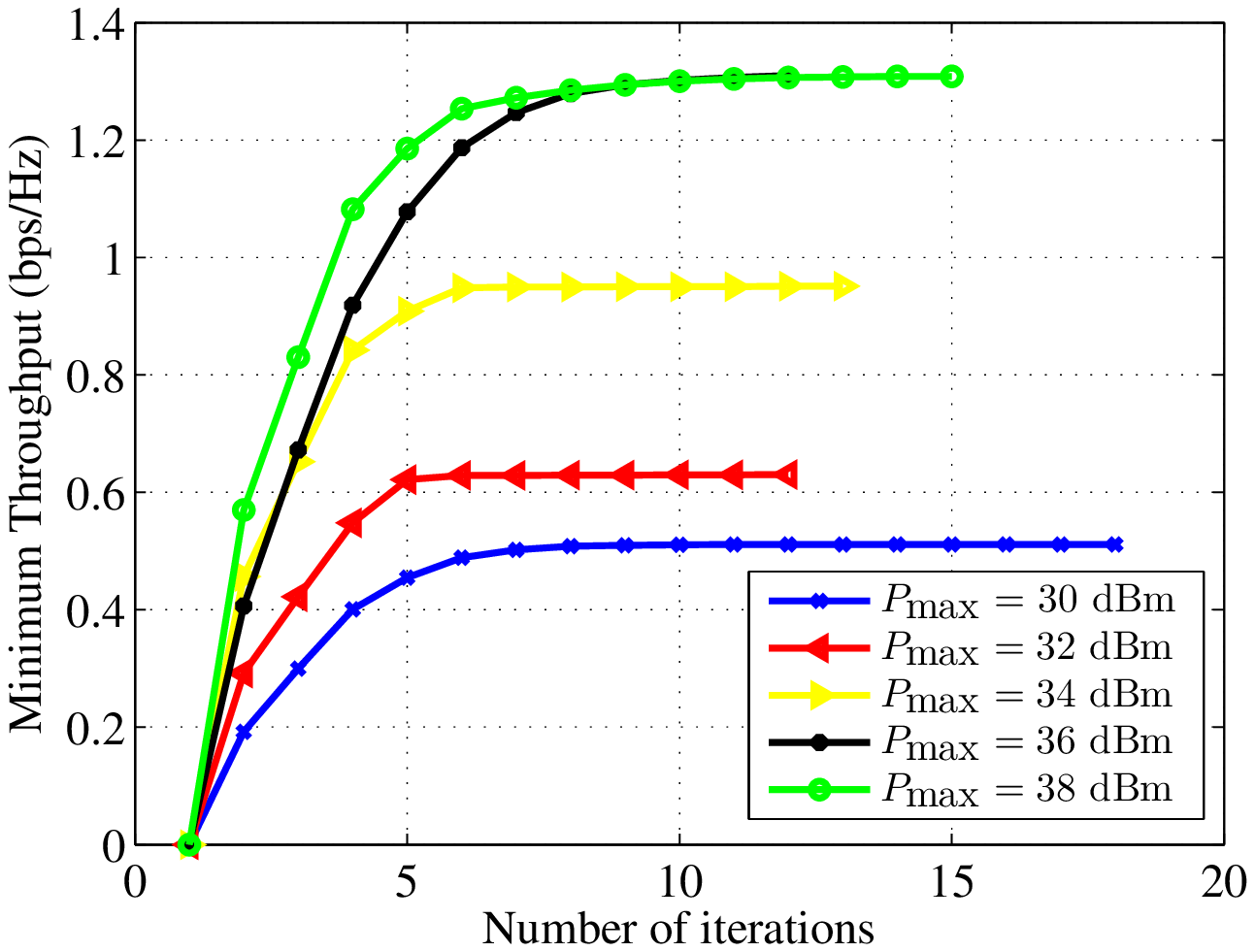}}
\hfill
\caption{The values of the objective function versus number of iterations for $\mathcal{K}=5$: (a) sum throughput (the objective in \eqref{form3}) in the sum throughput optimization problem, (b) minimum throughput (the objective in \eqref{maxmin1}) in the max-min throughput optimization problem.}
\label{httttt0}
\end{figure}

The convergence of the proposed algorithm for sum throughput optimization is shown in Fig. \ref{httttt0}.a via considering the values of the objective function in \eqref{form3} versus outer iterations number (see Table \ref{table:method}). It is observed that the objective values have a monotonic ascent property, as expected. In addition, the sum throughput is monotonically increasing w.r.t. the maximum transmit power $p_{\textrm{max}}$ in the first phase until around $p_{\textrm{max}} = 39$ dBm. After this value, increasing $p_{\textrm{max}}$ does not increase the sum throughput; this can be explained using the fact that each ER has a finite capacity energy storage $E_{{\textrm{max}}}$. Fig. \ref{httttt0}.b illustrates the objective values for the maxmin problem in \eqref{maxmin1}. Behaviors similar to those of the Fig. \ref{httttt0}.a can be seen. In this example, for the max-min problem, the amount of power in which saturation occurs ($p_{\textrm{max}} = 36$ dBm) is less than that of the sum throughput problem; because in the max-min problem, the optimal WET time $\tau_{\textrm{opt}}$ is greater (to be shown in the next subsection).

\subsection{The sum versus minimum throughput maximization}

\begin{figure}
\centering
\includegraphics[scale=.6]{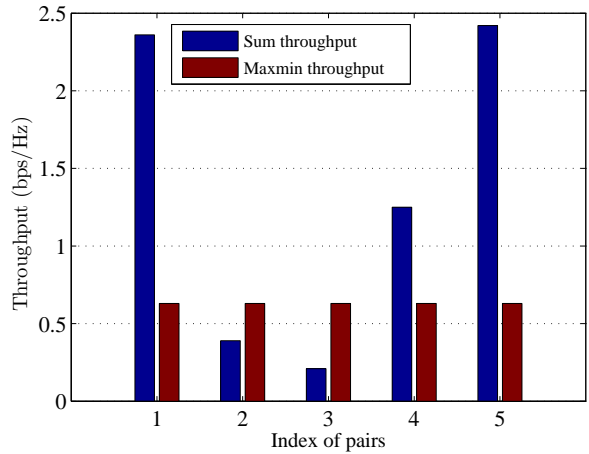}
\caption{An illustration of the values of the throughput for each pair in the sum and max-min throughput optimizations.}
\label{PA4}
\centering
\end{figure}
\begin{figure}
\centering
\includegraphics[scale=.6]{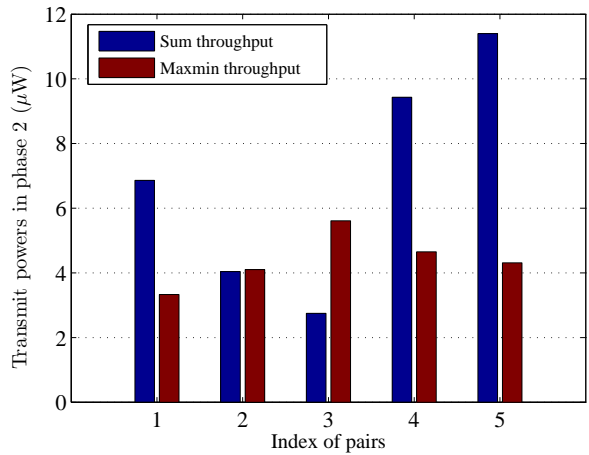}
\caption{An illustration of transmit powers in phase 2 in the sum and max-min throughput optimizations.}
\label{PA3}
\centering
\end{figure}

Next, we plot the sum and the minimum throughput in Fig \ref{PA4}. From this figure, it is observed that in the sum throughput maximization, the values of the maximum and minimum throughput are equal to $R_5=2.42$  $\textrm{bps}/\textrm{Hz}$ and $R_3=0.21$ $\textrm{bps}/ \textrm{Hz}$, respectively, which illustrates and emphasizes the unfair throughput allocation between pairs. Note that in the max-min design, we have $R_1 \simeq...\simeq R_5 \simeq 0.63$ $\textrm{bps}/ \textrm{Hz}$ to ensure the fairness among all pairs. However, the sum throughput of all pairs are equal to $6.63$ $\textrm{bps}/ \textrm{Hz}$ and $3.15$ $\textrm{bps}/ \textrm{Hz}$ for the sum and max-min designs, respectively. This shows that in the max-min throughput optimization, the sum throughput compromised for achieving the fairness.

Fig. \ref{PA3} illustrates transmit powers in phase 2 in both design problems. It is observed that the $3^{rd}$ IT transmits the most power among all ITs in the max-min throughput optimization problem. Also, as shown in Fig. \ref{finalom}, the $3^{rd}$ IT is located in the middle of the other pairs; and hence, this IT receives the highest level of interference at phase 2. Therefore, in phase 2, the $3^{rd}$ pair receives the most (harmful) interference that is generated by other pairs. Moreover, we know that in the max-min throughput optimization, the throughput allocation among all pairs should be fair. Consequently, for compensating the high interference level that is received by the $3^{rd}$ ET, the $3^{rd}$ IT should transmit a higher level of power in phase 2. We also show the time allocation for optimal values of $\mathbf{x}$ and ${\mathbf{p}}$. Herein, $\tau_{opt} = 0.22 $ and $\tau_{opt} = 0.47 $ for sum and minimum throughput optimizations, respectively.

\subsection{The improvement due to design of the energy waveform $\mathbf{x}$}
\begin{figure}
\hfill
\subfigure[]{\includegraphics[scale=.45]{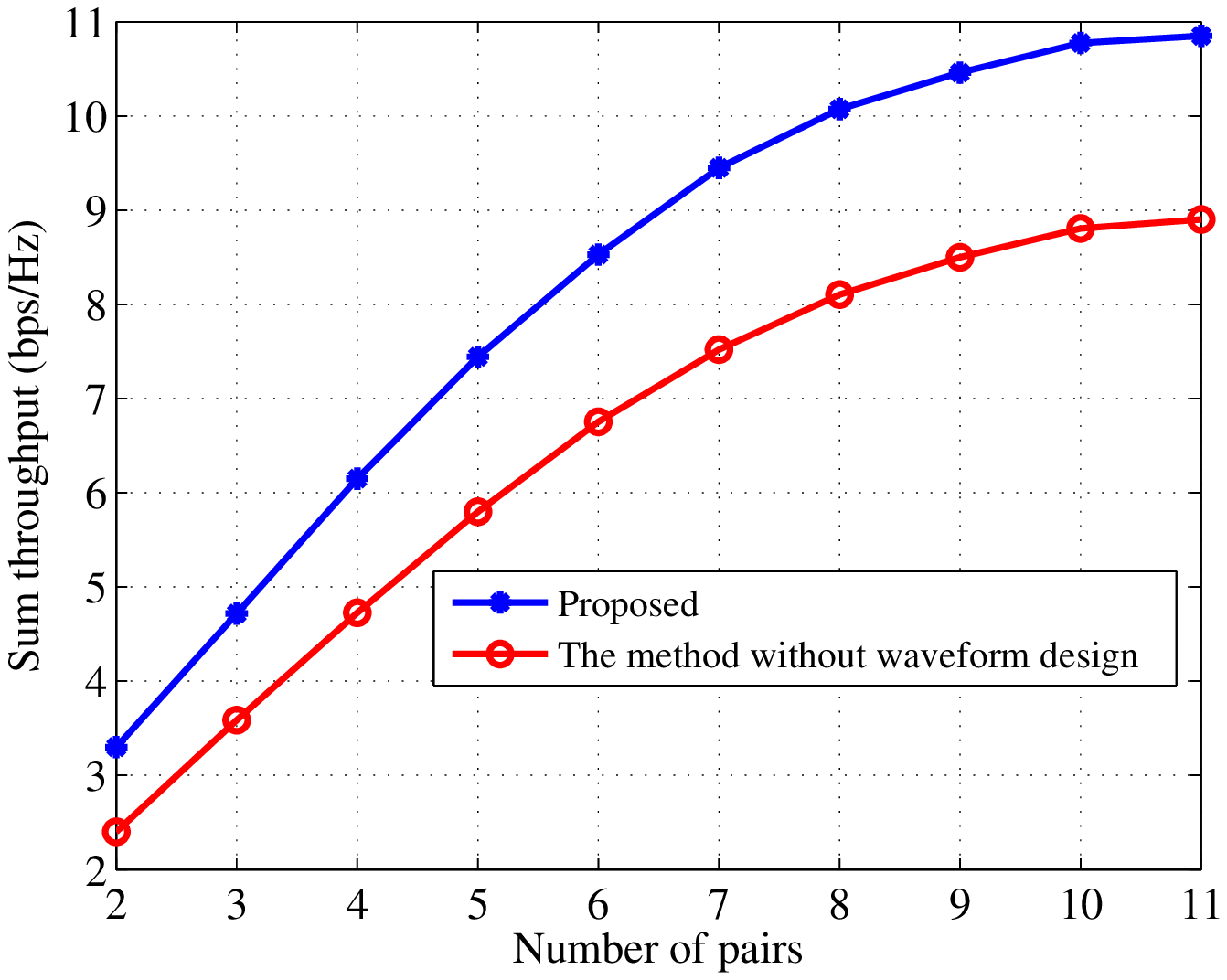}}
\hfill
\subfigure[]{\includegraphics[scale=.45]{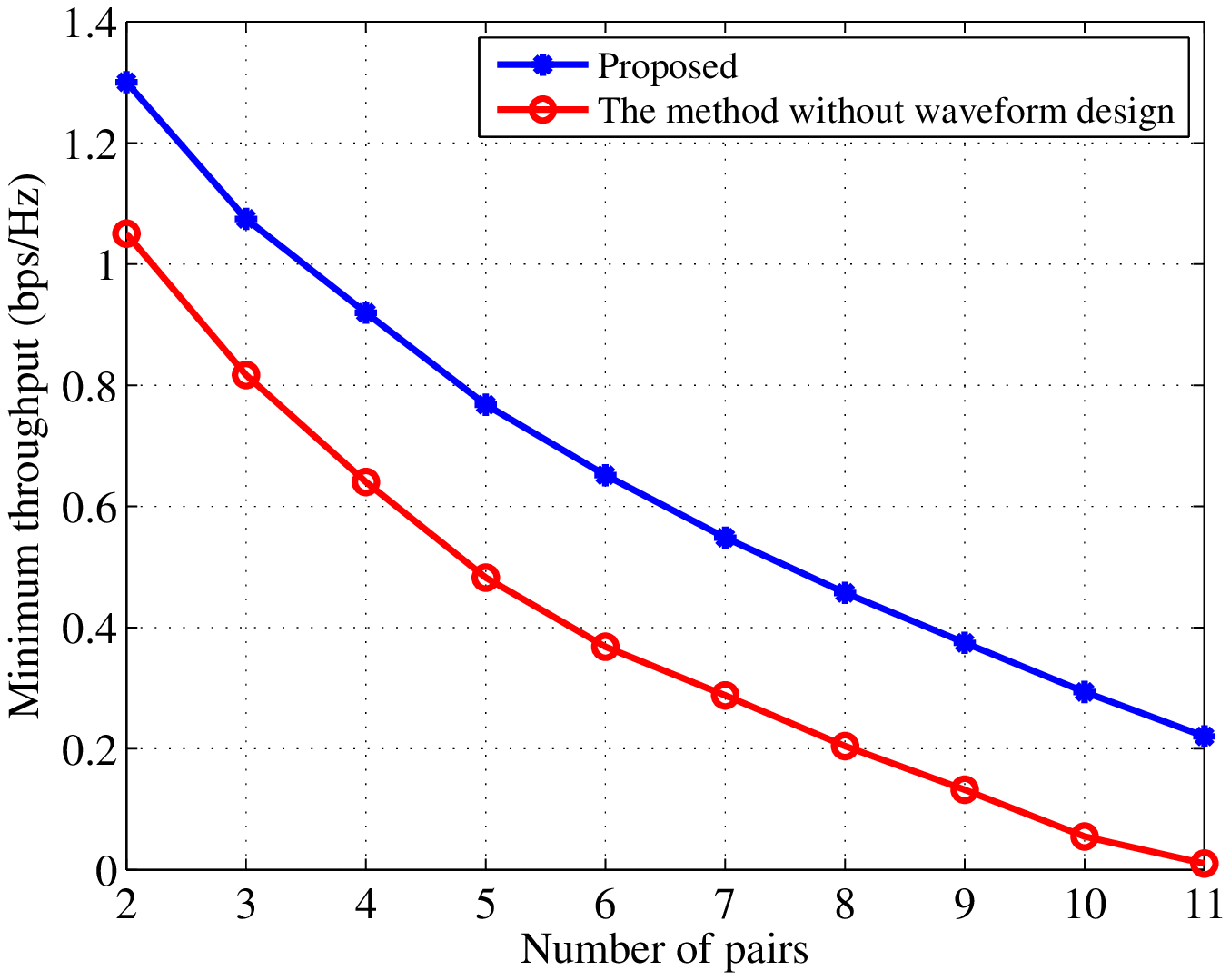}}
\hfill
\caption{Throughput values versus number of pairs: (a) sum throughput in the sum throughput maximization, (b) minimum throughput in the max-min optimization.}
\label{httttttt0}
\end{figure}
In this subsection, we investigate the sum and minimum throughput of pairs versus number of pairs ($\mathcal{K}$) to illustrate the improvement due to designing the energy waveform $\mathbf{x}$. We set $l_d=100$ (m) and as mentioned earlier, the distance between any pair and its adjacent pair will be equal to $\frac{l_d}{\mathcal{K}-1}$. Fig. \ref{httttttt0}.a and Fig. \ref{httttttt0}.b illustrate the sum and minimum throughput versus $\mathcal{K}$ in sum throughput maximization and max-min throughput optimization problems, respectively. It can be seen in both figures that the proposed resource allocation scheme (with the energy waveform design) achieves higher throughput than the scheme without energy waveform design for both cases. Note that for the method without joint energy waveform design, only the powers of the transmit energy signals are optimized,
i.e., ${\vert x_k \vert}^2$; indeed, in this case the degrees of freedom for phase $\textrm{arg}(x_k)$ are not exploited.
Also, Fig. \ref{httttttt0}.b shows that increasing the number of pairs decreases the minimum throughput in the max-min throughput optimization problem case. This behavior can be explained intuitively as follows: limiting a larger number of pairs in $l_d$ (m) results in higher interference level and thus lower minimum throughput. This phenomenon is also responsible for the saturation of curves in Fig. \ref{httttttt0}.a for the considered interval of $\mathcal{K}$.

\subsection{Non-linearity of EH circuit }
\begin{figure}
\centering
\includegraphics[scale=.55]{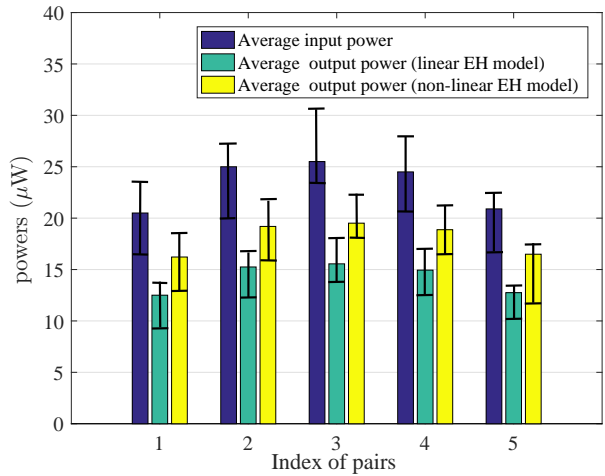}
\caption{An illustration of input and output powers of the EH circuits for linear versus non-linear EH models.}
\label{P23}
\centering
\end{figure}

Fig. \ref{P23} illustrates the input and output power levels of the EH circuit for linear and non-linear characteristics. Herein, we consider the results of Fig. \ref{P22} for the linear and non-linear models. Note that the input power range is around 20 $\mu$W; and as it can be seen from Fig. \ref{P22}, the rectifier works in the linear region. Therefore, as expected, the behavior and output powers assuming linear and non-linear models have minor differences. However, if one considers a scenario in which the rectifier goes to saturation, the corresponding results will be substantially different. Note also that the input/output powers vary for various channel realizations. Therefore, the results in Fig. \ref{P23} are reported in average sense but this figure also shows the aforementioned variations (by black signs) that belong to $[0.8 \bar{m} , 1.2 \bar{m}]$ with $\bar{m}$ being the average value.

\subsection{Effect of channel estimation error}

\begin{figure}
\hfill
\subfigure[]{\includegraphics[scale=.35]{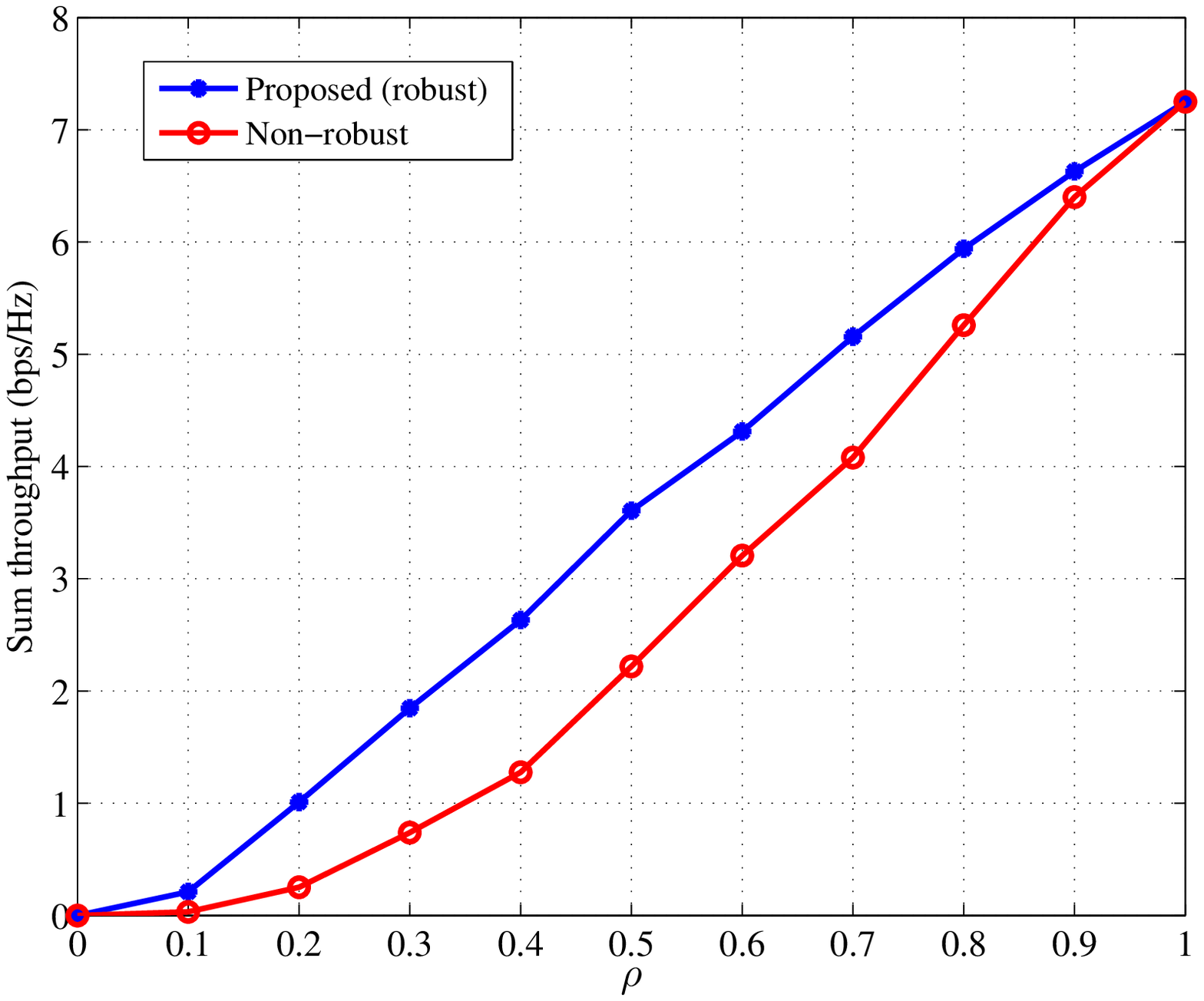}}
\hfill
\subfigure[]{\includegraphics[scale=.36]{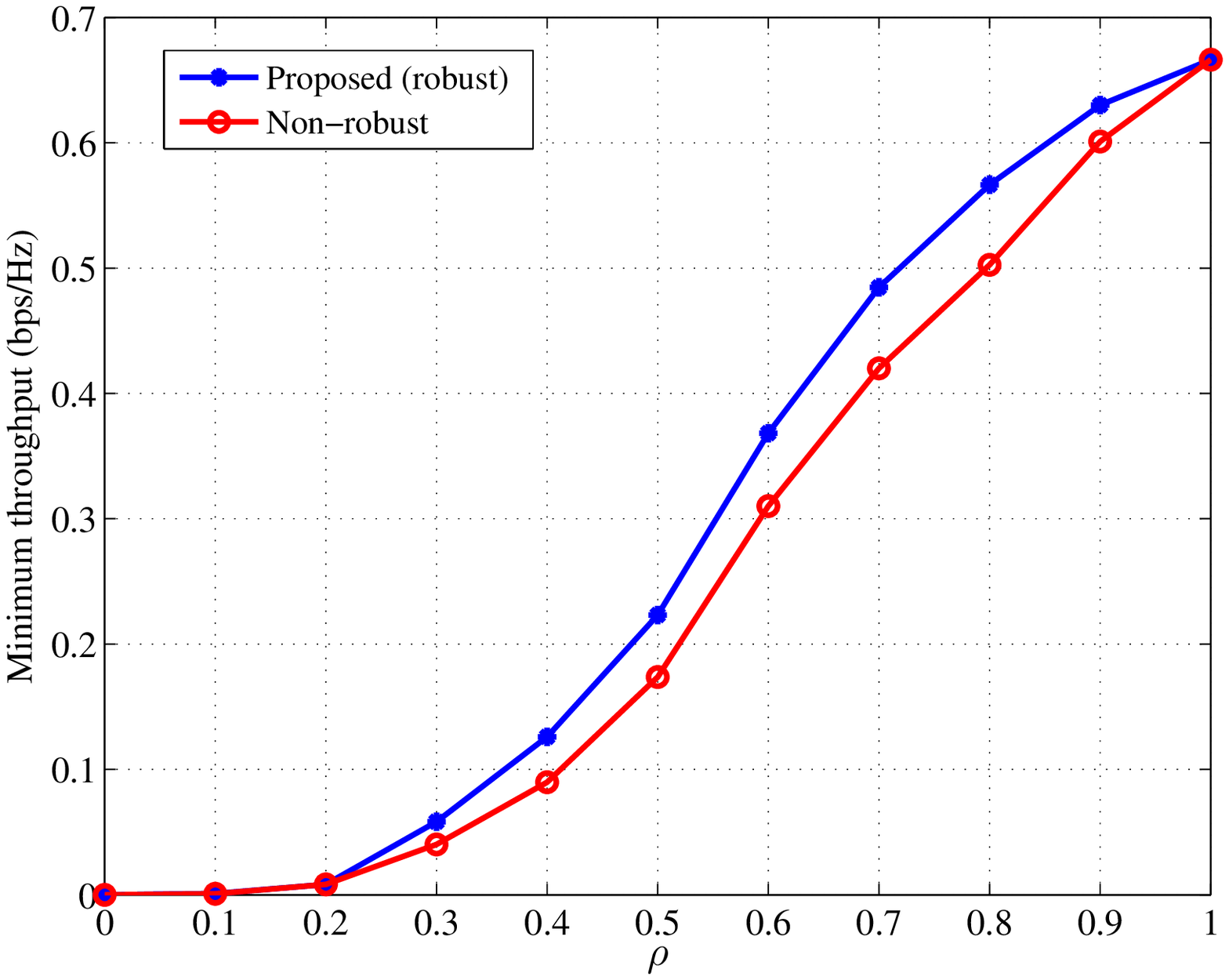}}
\hfill
\caption{Throughputs versus the estimation quality $\rho$ for $\mathcal{K}=5$: (a) sum throughput in the sum throughput maximization, (b) minimum throughput in the max-min optimization.}
\label{gtgt}
\end{figure}

\begin{figure}
\centering
\includegraphics[scale=.4]{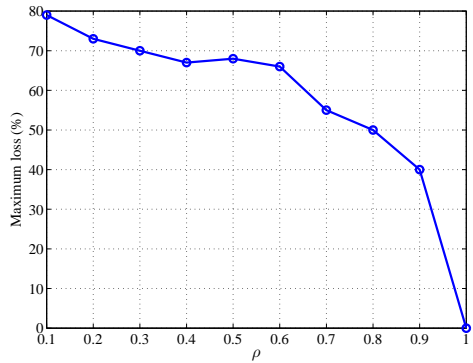}
\caption{The loss $\xi (\rho) $ versus the estimation quality $\rho$ in the max-min throughput optimization.}
\label{PA2}
\centering
\end{figure}
Further, we investigate the effect of the channel estimation error. The proposed method considers the channel estimation errors in the design stage and thus can be considered \textit{robust} w.r.t. errors. We compare the throughputs of the devised method with those of \textit{non-robust} method. The non-robust method herein refers to the case in which estimation errors are not taken into account in the design stage, i.e., assuming the channel estimates are accurate and error-free. The throughput comparisons in average sense over 100 random channel realizations for both sum and minimum throughput maximization are depicted in Fig. \ref{gtgt}.a and Fig. \ref{gtgt}.b, respectively. As expected, the sum and minimum throughput values in sum and minimum throughput designs increase with the estimation quality $\rho$. In addition, performance gains of the proposed method over the non-robust method  for both sum and minimum throughputs can be seen. The performance gain is not significant specially for max-min design; however, there might be  some channel realizations in which adopting the non-robust scheme instead of the robust one will cause a considerable performance degradation. Therefore, we define the loss parameter as
\begin{equation}  \label{a100}
\xi (\rho) \triangleq 1- \frac{R_{nr}(\rho)}{R_{r}(\rho) },
\end{equation}
where $R_{nr}(\rho)$ and $R_{r}(\rho)$ denote the throughputs for the non-robust and the robust schemes, respectively. Fig. \ref{PA2} illustrate the maximum value of $\xi (\rho)$ for maxmin design considering 100 random realizations of
channel coefficients. It is observed that increasing $\rho$, decreases the loss as expected. In addition, adopting the robust scheme results in significantly higher minimum throughput values in the max-min optimization case.

\subsection{Asymmetric scenario} \label{asym}
\begin{figure}
\centering
\begin{tikzpicture}[even odd rule,rounded corners=2pt,x=12pt,y=12pt,scale=.8]
\draw[thick,fill=green!10] (-15,5) rectangle ++(2.5,2) node[midway]{$\textrm{ET}_1$};
\draw[thick] (-12.5,6)--++(1.5,0)--+(0,1);
\draw[thick,fill=gray!15] (-11,7)--++(.75,0.5)--++(-1.5,0)--++(.75,-.5)--++(0,-.1);
\draw[thick,fill=green!10] (-15,-3.5) rectangle ++(2.5,2) node[midway]{$\textrm{ET}_2$};
\draw[thick] (-12.5,-2.5)--++(1.5,0)--+(0,1);
\draw[thick,fill=gray!15] (-11,-1.5)--++(.75,0.5)--++(-1.5,0)--++(.75,-.5)--++(0,-.1);
\draw[-,dashed,line width=.2mm,black!100](-11,7.5)--++(6.5,0);
\draw[-,dashed,line width=.2mm,black!100](-11,-1)--++(6.5,0);
\node [] at (-7.55,-.3){10 (m)};
\draw[-,dashed,line width=.2mm,black!100](-11,-1)--++(0,7);
\draw[<->,dashed,line width=.2mm,black!100](-4.5,-1)--++(0,8.5);
\node [] at (-2.7,3.25){12 (m)};
\draw[->,line width=.5mm,black!100](-4.5,7.5)--++(0,6.5);
\draw[->,line width=.5mm,black!100](-4.5,7.5)--++(6.5,0);
\draw [<->,line width=.4mm] (-2.5,7.5) arc (0:45:25pt);
\node [] at (-1.7,8.4){45};
\draw[<->,dashed,line width=.2mm,black!100](-4.5,7.5)--++(4.5,4.5)node[pos=0.5,above,sloped]{${\Delta} \textrm{x}$};
\draw[thick,fill=red!10] (2,10) rectangle ++(2.5,2) node[midway]{ $\textrm{IT}_1$};
\draw[thick] (2,11)--++(-1.5,0)--+(0,1);
\draw[thick,fill=gray!15] (.5,12)--++(.75,0.5)--++(-1.5,0)--++(.75,-.5)--++(0,-.1);
\draw[thick,fill=red!10] (-3,-3.5) rectangle ++(2.5,2) node[midway]{ $\textrm{IT}_2$};
\draw[thick] (-3,-2.5)--++(-1.5,0)--+(0,1);
\draw[thick,fill=gray!15] (-4.5,-1.5)--++(.75,0.5)--++(-1.5,0)--++(.75,-.5)--++(0,-.1);
\end{tikzpicture}
\caption{The geometry for the considered asymmetric scenario.}
\label{farIT}
\centering
\end{figure}
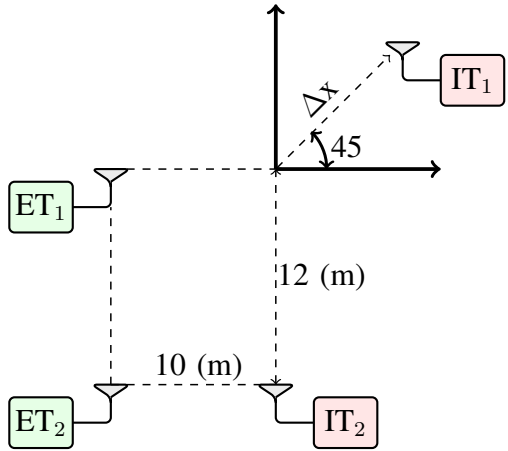

\begin{figure}
\centering
\includegraphics[scale=.45]{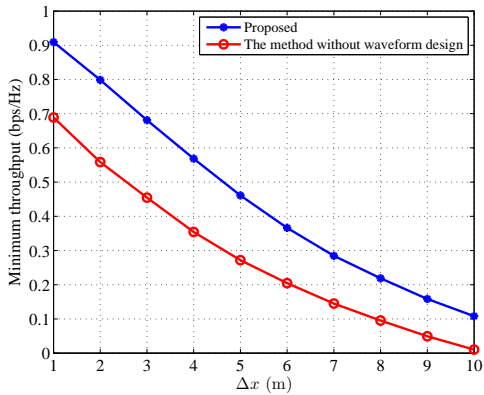}
\caption{Minimum throughput versus distance parameter $\Delta x$ (refer to Fig. \ref{farIT}) in max-min design for the asymmetric scenario.}
\label{ETIT}
\centering
\end{figure}

Finally, we consider an asymmetric geometry for pairs in numerical examples. For the purpose of
evaluating the performance of the proposed methods in a more general situation, we consider the case that $ \textrm{IT}_1$ gets far away in $y=x$ axis and other nodes are fixed as shown in Fig. \ref{farIT} (assuming $\mathcal{K}=2$). The general observations for the convergence, fairness, etc. are similar to those for the symmetric scenario. We now investigate the minimum throughput associated with the setup above (i.e., the throughput of the $1^{st}$ pair) which is shown in Fig. \ref{ETIT} for the case of max-min design. It is observed that moving away $ \textrm{IT}_1$ results in decreasing the minimum throughput. This can be explained by noting the fact that the received interference from $ \textrm{IT}_2$ to $ \textrm{ET}_1$ does not change (because $d_{2,2}$ and $d_{2,1}$ remain unchanged). Furthermore, the power of the received desired signal from $ \textrm{IT}_1$ to $ \textrm{ET}_1$ is decreased when increasing $d_{1,1}$ and $d_{1,2}$.

\section{Conclusion} \label{con}
In this paper, we studied the wireless powered $\mathcal{K}$-user IFC with the harvest-then-transmit protocol, consisting of $\mathcal{K}$ ET-IT pairs. 
The joint optimization of the energy waveform in first phase, the transmit powers in second phase, and the time allocation over the two phases was considered in order to maximize the sum throughput as well as the minimum throughput of the network. The aforementioned design problems were non-convex and thus difficult to be solved optimally. Therefore, we proposed a method based on alternating projections and MaMi techniques. Applying the proposed methods to the design problems provides stationary points of the problems (under some mild conditions). The effectiveness of the proposed methods was illustrated by numerical examples in various scenarios. The importance of designing the collaborative energy waveform and considering the channel estimation error along with non-linearity in EH circuit were also exemplified. The fairness in the network was also shown by comparing the throughput of users for sum and min throughput optimizations.

\appendices
\section{A selection of $\beta$ in \eqref{jj}} \label{app1}
The value of $\beta$ should be selected such that $E^{nl}_k (\mathbf{x}) + \beta
 \mathbf{x} {\mathbf{x}}^{H}$ is convex w.r.t. $\mathbf{x}$. Indeed, we should have ${\nabla}^{2} E^{nl}_k (\mathbf{x}) + \beta \mathbf{I}_{\mathcal{K}} \succeq \mathbf{0}$. It is verified that
\begin{equation*}
{\nabla}^{2} {E^{nl}_k (\mathbf{x})}  = \widetilde{\alpha}_{k} \mathbf{Q}_{k}    + \widetilde{\gamma}_{k} \mathbf{Q}_{k} \mathbf{x}  {\mathbf{x}}^{H}   \mathbf{Q}_{k},
\end{equation*}
with
\begin{equation}
\widetilde{\alpha}_{k}=\frac{2\tau N_k \widetilde{a}_k}{1-\Omega_k} \left( \textrm{exp} \left(-2 \left( \widetilde{a}_k   {\mathbf{x}}^{H}   \mathbf{Q}_{k}  \mathbf{x}   + \widetilde{a}_k \widetilde{b}_k \right) \right)
 + \textrm{exp} \left(- \widetilde{a}_k   {\mathbf{x}}^{H}   \mathbf{Q}_{k}  \mathbf{x}   +\widetilde{a}_k \widetilde{b}_k \right)
 \right)  \geq 0,
 \end{equation}
 \begin{equation} \label{apf}
    \widetilde{\gamma}_{k} =\frac{2\tau N_k \widetilde{a}_k}{1-\Omega_k}  \left( 2\widetilde{a}_k \left( \textrm{exp} \left(-2 \left( \widetilde{a}_k   {\mathbf{x}}^{H}   \mathbf{Q}_{k}  \mathbf{x}   + \widetilde{a}_k \widetilde{b}_k \right) \right)
 - \textrm{exp} \left(- \widetilde{a}_k   {\mathbf{x}}^{H}   \mathbf{Q}_{k}  \mathbf{x}   +\widetilde{a}_k \widetilde{b}_k \right)  \right)
 \right),
\end{equation}
and noting that ${\mathbf{Q}}_{k} \triangleq  {{\mathbf{h}}}_{k} {{{\mathbf{h}}}}^{H}_{k} \succeq \mathbf{0}$. As $\widetilde{\alpha}_{k} \geq 0$, it suffices to choose $\beta$ such that $\widetilde{\gamma}_{k} \mathbf{Q}_{k} \mathbf{x}  {\mathbf{x}}^{H}   \mathbf{Q}_{k} + \beta \mathbf{I}_{\mathcal{K}} \succeq \mathbf{0}$; or equivalently,
\begin{equation}\label{appp}
\beta \mathbf{I}_{\mathcal{K}} \succeq - \widetilde{\gamma}_{k} \mathbf{Q}_{k} \mathbf{x}  {\mathbf{x}}^{H}   \mathbf{Q}_{k}.
\end{equation}
Considering the fact that $ \mathbf{Q}_{k} \mathbf{x}  {\mathbf{x}}^{H}   \mathbf{Q}_{k}$ is a rank-1 matrix, the inequality in \eqref{appp} holds if
\begin{equation} \label{66}
\beta \geq -\widetilde{\gamma}_{k} {\mathbf{x}}^{H}  \mathbf{Q}^{2}_{k} \mathbf{x}.
\end{equation}
Next, it is worth noting that by \eqref{apf} we can write
\begin{equation} \label{55}
 -\widetilde{\gamma}_{k}  \leq \frac{4\tau N_k {\widetilde{a}_k}^{2}}{1-\Omega_k}  \textrm{exp} \left(\widetilde{a}_k \widetilde{b}_k \right),
\end{equation}
and also
 \begin{equation} \label{56}
{\mathbf{x}}^{H}  \mathbf{Q}^{2}_{k} \mathbf{x} \leq {\| \mathbf{x} \|}^{2}_{2} \mathbf{\lambda}_{\textrm{max}} \left( {\mathbf{Q}}^{2}_{k} \right).
\end{equation}
Finally, using \eqref{66}, \eqref{55}, \eqref{56}, and considering the constraint ${\| \mathbf{x} \|}^{2}_{2} \leq \sum_{k=1}^{\mathcal{K}} {p}_{\textrm{max},{k}}$, we can select $\beta > \beta_0$ with
\begin{equation}
\beta_0 =\left( \frac{4\tau N_k {\widetilde{a}_k}^{2}}{1-\Omega_k}  \textrm{exp} \left(\widetilde{a}_k \widetilde{b}_k \right) \right) \left( \lambda_{\textrm{max}} \left( {\mathbf{Q}}^{2}_{k} \right) \right) \left( \sum_{k=1}^{\mathcal{K}} {p}_{\textrm{max},{k}} \right).
\end{equation}

\bibliographystyle{IEEETran}
\bibliography{myreff}
\end{document}